\documentclass[%
reprint,
superscriptaddress,
preprintnumbers,
nofootinbib,
 amsmath,amssymb,
 aps,prd
]{revtex4-1}

\usepackage{graphicx}
\usepackage[dvipsnames]{xcolor}
\usepackage{dcolumn}
\usepackage{bm}
\usepackage[colorlinks]{hyperref}
\usepackage{chemformula}
\usepackage[utf8]{inputenc}

\newcommand\myshade{80}
\colorlet{mylinkcolor}{ForestGreen}
\colorlet{mycitecolor}{Red}
\colorlet{myurlcolor}{violet}
\newcommand{\DA}{\bar{\mathcal{D}}}
\newcommand{\vect}{\boldsymbol}

\hypersetup{
  linkcolor  = mylinkcolor!\myshade!black,
  citecolor  = mycitecolor!\myshade!black,
  urlcolor   = myurlcolor!\myshade!black,
  colorlinks = true
}

\newcommand{\GRAPPA}{%
Gravitation Astroparticle Physics Amsterdam (GRAPPA),\\
Institute for Theoretical Physics Amsterdam
and Delta Institute for Theoretical Physics,\\
University of Amsterdam, Science Park 904, 1098 XH Amsterdam, The Netherlands}
\newcommand{\OKC}{%
The Oskar Klein Centre for Cosmoparticle Physics, Department of
Physics, Stockholm University, Alba Nova, 10691 Stockholm, Sweden}
\newcommand{\Nordita}{%
Nordita, KTH Royal Institute of Technology and Stockholm University, Roslagstullsbacken 23, 10691 Stockholm, Sweden}
\newcommand{\NCBJ}{National Centre for Nuclear Research, 05-400 Otwock, \'{S}wierk, Poland}

\newcommand{\dbd}[2]{\frac{\mathrm{d}#1}{\mathrm{d}#2}}

\begin{document}

\preprint{NORDITA-2018-119; LCTP-18-26}

\title{Digging for Dark Matter:\\ 
Spectral Analysis and Discovery Potential of Paleo-Detectors}

\author{Thomas D. P. Edwards}
\email{t.d.p.edwards@uva.nl}
\affiliation{\GRAPPA}
 
\author{Bradley J. Kavanagh}
\email{b.j.kavanagh@uva.nl}
\affiliation{\GRAPPA}

 \author{Christoph Weniger}
 \email{c.weniger@uva.nl}
\affiliation{\GRAPPA}
 
\author{\\Sebastian Baum}
 \email{sbaum@fysik.su.se}
 \affiliation{\OKC}
 \affiliation{\Nordita}
 
\author{Andrzej~K.~Drukier}
\email{adrukier@gmail.com}
\affiliation{\OKC}

\author{Katherine Freese}
\email{ktfreese@umich.edu}
\affiliation{\OKC}
\affiliation{\Nordita}
\affiliation{Leinweber Center for Theoretical Physics, University of Michigan, Ann Arbor, MI 48109, USA}

\author{Maciej~G\'{o}rski}
\email{maciej.gorski@ncbj.gov.pl}
\affiliation{\NCBJ}

\author{Patrick Stengel}
\email{patrick.stengel@fysik.su.se}
\affiliation{\OKC}





\date{\today}

          
\begin{abstract}
   Paleo-detectors are a recently proposed method for the direct detection of Dark Matter (DM). In such detectors, one would search for the persistent damage features left by DM--nucleus interactions in ancient minerals. Initial sensitivity projections have shown that paleo-detectors could probe much of the remaining Weakly Interacting Massive Particle (WIMP) parameter space. In this paper, we improve upon the cut-and-count approach previously used to estimate the sensitivity by performing a full spectral analysis of the background- and DM-induced signal spectra. We consider two scenarios for the systematic errors on the background spectra: i) systematic errors on the normalization only, and ii) systematic errors on the shape of the backgrounds. We find that the projected sensitivity is rather robust to imperfect knowledge of the backgrounds. Finally, we study how well the parameters of the true WIMP model could be reconstructed in the hypothetical case of a WIMP discovery.
\end{abstract}

\maketitle

\section{Introduction}

Direct detection experiments search for Weakly Interacting Massive Particle (WIMP) Dark Matter (DM) by looking for rare, low-energy nuclear recoils induced by the scattering of DM particles off a target's nuclei~\cite{Drukier:1983gj,Goodman:1984dc,Drukier:1986tm,Spergel:1987kx}. As yet, no conclusive evidence for DM has been reported by direct detection experiments.  The leading upper limits on DM--nucleus interactions for WIMPs with masses greater than $\sim10\,$GeV stem from ton-scale liquid noble gas experiments~\cite{Akerib:2016vxi,Cui:2017nnn,Amaudruz:2017ekt,Aprile:2018dbl,Agnes:2018fwg}. Meanwhile, cryogenic bolometric detectors have started to probe DM--nucleus interactions for lighter DM candidates~\cite{Hehn:2016nll,Agnese:2017jvy,Angloher:2017sxg,Petricca:2017zdp,Agnese:2018gze}. The noteworthy exception is DAMA, which has been reporting evidence~\cite{Bernabei:2008yi,Bernabei:2013xsa,Bernabei:2018yyw} for an annually modulated signal~\cite{Drukier:1986tm,Freese:2012xd} compatible with WIMP DM~\cite{Gondolo:2005hh,Petriello:2008jj,Chang:2008xa,Fairbairn:2008gz,Savage:2008er,Bottino:2008mf,Baum:2018ekm,Kang:2018qvz,Herrero-Garcia:2018lga} for more than a decade, although this claim is in tension with upper limits from other direct detection experiments~\cite{Savage:2008er,Savage:2009mk,Savage:2010tg,McCabe:2011sr,Catena:2016hoj}.

Progress in the search for DM heavier than $\sim 10\,$GeV relies on maximizing the exposure (the product of target mass and integration time) of the experiment. Instead, to probe ever lighter DM, experiments must achieve sensitivity to lower and lower nuclear recoil energies.  In addition, both mass regimes require exquisite control of a variety of possible background sources, from cosmic rays to intrinsic radioactivity. A number of experiments with lower energy thresholds, larger exposures or, ideally, both will continue the search for lighter and more weakly-interacting DM in the next 5--10 years~\cite{EURECA,Agnese:2016cpb,Aalbers:2016jon,XENONNT,Aalseth:2017fik,Akerib:2018lyp}.

Recently, Refs.~\cite{Baum:2018tfw,Baum:2018zzz} proposed {\it paleo-detectors} for the direct detection of DM\footnote{For brevity, we use the term `Dark Matter' (DM) in this work, but will be considering specifically WIMP DM throughout.}. In paleo-detectors, one would examine ancient minerals extracted from $\mathcal{O}(10)\,$km below the surface of the Earth for traces of DM interactions with atomic nuclei. In lieu of the multi-ton target masses used in conventional direct detection experiments, paleo-detectors take advantage of the fact that DM may have been interacting with the target material for as long as $\sim 1\,$Gyr. In certain minerals, including those long used as Solid State Track Detectors (SSTDs)~\cite{Fleischer:1964,Fleischer383,Fleischer:1965yv,GUO2012233}, DM-induced nuclear recoils would give rise to $1-500\,$nm long damage tracks. In many materials, such damage tracks would be preserved over timescales much longer than 1\,Gyr. Paleo-detectors could thus obtain much larger exposures than conventional detectors even if only a small amount of target material can be investigated. For example, reading out 100\,g of material which has been recording DM-induced events for 1\,Gyr provides an exposure which could only be matched in the laboratory by observing $10^{4}$\,tons of target mass for 10\,yr. Further, the relatively small target masses required for paleo-detectors can be obtained from depths much greater than those of  underground laboratories in which conventional direct detection experiments are usually operated, providing an unprecedented level of shielding from cosmic rays.

In conventional direct detection  experiments, nuclear recoils are detected by scintillation, ionization, and phonon signals in the detector~\cite{Undagoitia:2015gya}. In paleo-detectors, the observational signature would be nano-scale defects in the minerals. These may be observed through a variety of read-out methods such as X-ray microscopy~\cite{Kirz1985,XrayMicroscopy,Schaff:2015} or ion-beam microscopy~\cite{Hill:2012}.  Mineral-based searches, initially for monopoles and then for DM, have been proposed and performed before~\cite{PhysRevLett.52.1265,PhysRevLett.56.1226,Collar:1994mj,   PhysRevC.52.2216,SnowdenIfft:1995ke,SnowdenIfft:1995qw,SnowdenIfft:1997hd,Baltz:1997dw}. However, modern high-resolution imaging techniques~\cite{Stevens2008,Urban2009,deJonge2010,vanGastel:2012, Joens:2013,Rodriguez:2014,Holler:2014} as well as the availability of rocks from deeper underground may significantly improve the prospects for DM detection. In particular, measurements of nanometre-length tracks could provide sensitivity to recoil energies as low as 100\,eV.  A more detailed study of backgrounds, target minerals and read-out methods for paleo-detectors was recently presented in Ref.~\cite{Baum:2018zzz}.

In this work, we explore the prospects for excluding or discovering DM with paleo-detectors. In Refs.~\cite{Baum:2018tfw,Baum:2018zzz}, a simple cut-and-count analysis with a sliding signal window in track length was employed to estimate the sensitivity of paleo-detectors. Here, we adopt more sophisticated statistical techniques, making use of information contained in the spectral shape of the track length distributions. Using realistic  distributions for backgrounds induced by neutrinos \cite{OHare:2016pjy} and radioactivity (developed in Refs.~\cite{Baum:2018tfw,Baum:2018zzz}), this approach allows us to explore the impact of different systematic background uncertainties on projected sensitivities. Further, we examine how well the DM properties (mass and cross section) could be reconstructed by paleo-detectors in the case of a discovery.

As in Refs.~\cite{Baum:2018tfw,Baum:2018zzz} we consider two fiducial read-out scenarios. The first assumes a high track-length resolution but a relatively small exposure (referred to as \textit{high resolution} in the following). The second assumes worse track-length resolution, which should in principle allow more material to be analyzed, facilitating a larger exposure (referred to as \textit{high exposure}). The high resolution configuration is particularly well suited to probing DM with masses below $\sim 10\,$GeV while the high exposure configuration is geared more towards heavier DM.

A wide range of minerals are well suited to be paleo-detectors. As described in the discussion of mineral optimization in Ref.~\cite{Baum:2018zzz}, minerals can be broadly divided into classes suitable for different applications of paleo-detectors, based on their chemical composition. We consider 4 different minerals in this work, chosen to represent paleo-detectors suitable for probing spin-independent DM--nucleus interactions:
\begin{itemize}
\itemsep0em 
    \item halite - \ch{NaCl},
    \item olivine - \ch{Mg_{1.6}Fe^{2+}_{0.4}(SiO4)},
    \item sinjarite - \ch{CaCl2$\cdot$ 2 (H2O)},
    \item nchwaningite - \ch{Mn^{2+}2SiO3(OH)2$\cdot$(H2O)}.
\end{itemize}
These particular target materials are also selected for their low levels of natural radioactive contamination, helping to suppress radioactivity-induced backgrounds. Average uranium concentrations in the Earth's crust are a few parts per million (ppm) by weight, which would lead to unacceptably high levels of background due to radioactivity. Minerals formed in {\it ultra-basic rock} deposits, stemming from the Earth's mantle, are much more radiopure. Examples of such minerals investigated here are olivine and nchwaningite, for which we assume uranium concentrations of 0.1 parts per billion. Even less contaminated by radioactive elements may be minerals in {\it marine evaporite} deposits formed at the bottom of evaporating oceans. We assume uranium concentrations of 0.01 parts per billion for such minerals and use halite and sinjarite as examples. Halite and olivine are very common minerals. In contrast, sinjarite and nchwaningite are less abundant but contain hydrogen. Although hydrogen makes up only a small fraction of these minerals by mass, its presence plays an important role in reducing neutron-induced backgrounds, as we discuss later.

We note that Refs.~\cite{Baum:2018tfw,Baum:2018zzz} studied both halite and olivine, allowing a straightforward comparison with our results. Nchwaningite was also used in~\cite{Baum:2018zzz} and our results are comparable to nickelbischofite [\ch{NiCl$_2\cdot$ 6 (H2O)}], used in~\cite{Baum:2018tfw}, due to its similar chemical composition. Sinjarite gives similar results to epsomite [\ch{Mg(SO4)$\cdot$ 7 (H2O)}] used in~\cite{Baum:2018tfw,Baum:2018zzz}.

The rest of this paper is organized as follows. In Sec.~\ref{sec:Theory}, we discuss paleo-detectors in more detail, including the calculation of signal spectra, track lengths, and the most relevant background sources. In Sec.~\ref{sec:Sensitivity}, we present the projected upper limits and discovery reach for the minerals listed above. In Sec.~\ref{sec:Reconstruction}, we use benchmark-free techniques to determine contours in the (DM mass)--(cross section) plane which could be reconstructed with paleo-detectors in the hypothetical case of a future discovery. We discuss a number of challenges for paleo-detectors in Sec.~\ref{sec:Challenges}. Our conclusions are presented in Sec.~\ref{sec:Conclusion}.

Code for performing all calculations presented in this paper is publicly available \href{https://github.com/tedwards2412/paleo_detectors}{here}~\cite{paleocode}.

\section{Theory}
\label{sec:Theory}

\subsection{Signal from WIMP Scattering}

For elastic scattering of DM with mass $m_\chi$ off nuclei with mass $m_N$, the differential recoil rate as a function of recoil energy $E_R$ per unit target mass is given by~\cite{Cerdeno:2010jj}: 
\begin{align}
    \dbd{R}{E_R} = \frac{\rho_\chi}{m_N m_\chi} \int_{v_\mathrm{min}}^\infty v f(\mathbf{v}) \dbd{\sigma}{E_R}\,\mathrm{d}^3\mathbf{v}\,.
\end{align}
The integral is over DM velocities $\mathbf{v}$, with $v = |\mathbf{v}|$
and $v_\mathrm{min} = \sqrt{m_N E_R/2\mu_{\chi N}^2}$.
We assume standard spin-independent (SI) interactions, with equal couplings to protons and neutrons. In this case, the differential cross section can be re-written in terms of the DM--nucleon cross section at zero momentum transfer $\sigma_n^\mathrm{SI}$ as:
\begin{align}
    \dbd{\sigma}{E_R} = \sigma_{n}^\mathrm{SI}\frac{m_N }{2\mu_{\chi p}^2 v^2} A^2 F^2(E_R)\,.
\end{align}
Here, $\mu_{\chi N} \equiv m_\chi m_N/(m_\chi + m_N)$ is the reduced mass of the DM--nucleus system (and similarly for the DM-proton reduced mass $\mu_{\chi p}$). The factor of $A^2$ corresponds to the coherent enhancement for a nucleus composed of $A$ nucleons. The internal structure of the nucleus is encoded in the form factor $F^2(E_R)$, for which we assume the Helm parametrization~\cite{Helm:1956zz,Lewin:1995rx,Duda:2006uk}. The differential recoil rate then takes the standard form:
\begin{align}
\dbd{R}{E_R} = \frac{\rho_\chi \sigma_n^\mathrm{SI}}{2 m_\chi \mu_{\chi p}^2} A^2 F^2(E_R) \int_{v_\mathrm{min}}^\infty \frac{f(\mathbf{v})}{v}\,\mathrm{d}^3\mathbf{v}\,.
\end{align}

For the DM distribution, we assume the Standard Halo Model (SHM), fixing a benchmark value for the local density of $\rho_\chi = 0.3\,\mathrm{GeV}\,\mathrm{cm}^{-3}$~\cite{Green:2011bv,Green:2017odb} in order to compare directly with other direct detection experiments. However, we note that observational estimates of $\rho_\chi$ vary substantially~\cite{Read:2014qva}. In the SHM, the DM velocity distribution $f(\mathbf{v})$ follows a truncated Maxwell-Boltzmann distribution, for which we fix values for the Sun's speed $v_\odot= 248 \, \mathrm{km/s}$~\cite{Bovy:2012ba}, the local circular speed $v_\mathrm{c} = 235  \,
\mathrm{km/s}$~\cite{Koposov:2009hn}, and the Galactic escape speed
$v_\mathrm{esc} = 550\, \mathrm{km/s}$~\cite{Piffl:2013mla}. We do not consider here uncertainties on the speed distribution~\cite{Green:2017odb} or more recently suggested refinements to the SHM~\cite{Evans:2018bqy}.

Recoil spectra for DM and neutrino scattering are calculated using the publicly available \texttt{WIMpy} code~\cite{WIMpy-code}.

\subsection{Paleo-Detector Rates}

The rate of tracks produced with length $x_T$ per unit target mass is given by:
\begin{align}
    \dbd{R}{x_T} = \sum_{i}^\mathrm{nuclei} \xi_i \dbd{R_i}{E_R} \left(\dbd{E_R}{x_T}\right)_{i}\,,
\end{align}
where the index $i$ runs over the different target nuclei which make up the mineral. The rate of nuclear recoils (with initial energy $E_R$) per unit mass is given by $\mathrm{d}R_i/\mathrm{d}E_R$ and we weight by the mass fraction $\xi_i$ of each nucleus $i$. The track length as a function of initial recoil energy is calculated as:
\begin{align}
    x_T(E_R) = \int_0^{E_R} \left|\dbd{E}{x_T}\right|^{-1} \,\mathrm{d}E\,.
\end{align}
The stopping power $\mathrm{d}E/\mathrm{d}x_T$ as a function of energy must be calculated for each of the recoiling nuclei in a given target material. We use the publicly available SRIM package (\textbf{S}topping and \textbf{R}ange of \textbf{I}ons in \textbf{M}atter)~\cite{Ziegler1985,SRIM2010} to tabulate the stopping power, although analytic estimates are also possible~\cite{Wilson1977,Baum:2018zzz}. A more detailed discussion of the calculation of track lengths can be found in Ref.~\cite{Baum:2018zzz}.

In this paper we assume that recoiling hydrogen nuclei and $\alpha$-particles do not produce tracks which can be reconstructed. Whether such low-$Z$ tracks are observable will depend on the target material and read-out method and is a question which must be determined empirically. A discussion of this issue as well as a comparison of results with and without low-$Z$ tracks can be found in Ref.~\cite{Baum:2018zzz}.

The resolution at which track lengths can be measured depends on the read-out technique used. For a true track length of $x^\prime$, we assume that the measured track length $x$ is Gaussian distributed\footnote{This assumption will depend on the imaging technique used and in practice it may be necessary to quantify the probability distribution of the measured track length experimentally.} with track length resolution $\sigma_{x_T}$:
\begin{align}
P(x|x^\prime) = \frac{1}{\sqrt{2\pi \sigma_{x_T}^2}}\exp\left(-\frac{(x-x^\prime)^2}{2\sigma_{x_T}^2}\right)\,.
\end{align}

The number of tracks with lengths in the range $[x_a,\,x_b]$ is then:
\begin{align}
    N(x_a, x_b) = \epsilon \int_{0}^{\infty} W(x_T; x_a, x_b)\, \dbd{R}{x_T}(x_T)\,\mathrm{d}x_T\,,
\end{align}
where $\epsilon$ is the exposure, given by the product of the age of the mineral and the total mass of the sample analyzed. The window function $W$ captures resolution effects and is given by:
\begin{equation}
    W(x_T; x_a, x_b) = \frac{1}{2}\left[\mathrm{erf}\left(\frac{x_T - x_a}{\sqrt{2}\sigma_{x_T}}\right) - \mathrm{erf}\left(\frac{x_T - x_b}{\sqrt{2}\sigma_{x_T}}\right)\right]\,.
\end{equation}
We assume that the smallest measurable track length is $\sigma_{x_T}/2$ and consider tracks as long as $1000 \,\mathrm{nm}$. As we will see in Sec.~\ref{sec:Spectra}, DM-induced recoils do not typically induce tracks longer than this.

We will consider two scenarios for the analysis of paleo-detectors, as in Ref.~\cite{Baum:2018zzz}:
\begin{itemize}
    \item \textit{High resolution} - we assume a track length resolution of $\sigma_{x_T} = 1\,\mathrm{nm}$, which may be achievable with helium ion beam microscopy~\cite{Hill:2012} (using a focused ion beam~\cite{Lombardo:pp5019,Joens:2013} and/or pulsed lasers~\cite{ECHLIN20151,PFEIFENBERGER2017109,Randolph:2018} to remove layers of material which have already been imaged). In this case, we assume an exposure of $\epsilon = 0.01 \,\mathrm{kg}\,\mathrm{Myr}$, which would correspond to a few $\mathcal{O}(1 \,\mathrm{mm}^3)$ mineral samples, each with an age of 1 Gyr. 
    
    \item \textit{High exposure} - we assume a track length resolution of $\sigma_{x_T} = 15\,\mathrm{nm}$. Small angle X-ray scattering has been demonstrated to achieve such spatial resultions in three dimensions~\cite{Rodriguez:2014,Holler:2014,Schaff:2015}. However, such resolutions have not yet been demonstrated when imaging damage tracks arising from nuclear recoils. Here we assume an exposure of $\epsilon = 100 \,\mathrm{kg}\,\mathrm{Myr}$, corresponding to the analysis of larger samples of $\mathcal{O}(10 \,\mathrm{cm}^3)$.
\end{itemize}
This is not an exhaustive list of possible scenarios, see Ref.~\cite{Baum:2018zzz} for a discussion of a variety of read-out techniques.

We note that a number of techniques (both stratigraphic and radiometric) are used for dating rock samples~\cite{GTS2012}. Perhaps most relevant for $\mathcal{O}(\mathrm{Gyr})$-aged rocks is fission track dating, which should allow an age estimate which is accurate to $\sim 10\%$~\cite{Gallagher:1998,vandenHaute:1998}, though we will neglect dating uncertainties in our analysis\href{http://www.stratigraphy.org/ICSchart/ChronostratChart2018-08.jpg}{.}

Given the target materials we analyze, we note that DM candidates with $m_\chi \lesssim 500 \, \mathrm{MeV}$ do not give rise to a significant number of recoil tracks longer than $\sim 1\,\mathrm{nm}$, the best track length resolution assumed in our analysis. Thus, we only consider DM with $m_\chi \gtrsim 500 \, \mathrm{MeV}$, though it may be possible to probe lower mass DM with either better track length resolution or with target materials which allow for longer tracks.   

\subsection{Backgrounds}
\label{sec:Backgrounds}

Here, we summarize the most problematic backgrounds for DM searches with ancient minerals. While follow-up studies and direct calibration will likely lead to refined background modeling, we expect the estimates presented here to be representative. A more detailed discussion can be found in Ref.~\cite{Baum:2018zzz}.

We note that cosmogenic backgrounds should be negligible for materials obtained from a depth below $\sim 5 \,\mathrm{km}$, meaning that the dominant backgrounds will be neutrino interactions and intrinsic radioactive backgrounds in the target materials themselves.

\paragraph{Neutrinos ---}
Being weakly interacting particles, neutrinos represent an irreducible background to (non-directional) DM searches~\cite{Monroe:2007xp,Strigari:2009bq,Gutlein:2010tq,Billard:2013qya}. Neutrinos with MeV-energies can produce keV-scale nuclear recoils, thereby mimicking DM signatures. We calculate the neutrino-nucleus scattering rate following Ref.~\cite{Billard:2013qya} (and references therein). In our analysis, we include solar, atmospheric and diffuse supernova background (DSNB) neutrino  fluxes, with spectra and normalizations as compiled in Ref.~\cite{OHare:2016pjy}. Uncertainties on the present day neutrino fluxes vary substantially, from $\mathcal{O}(1\,\%)$ for $^8\mathrm{B}$ and hep neutrinos~\cite{Agostini:2017cav} to as much as $50\,\%$ for DSNB neutrinos~\cite{Beacom:2010kk}. Paleo-detectors probe neutrino fluxes over the past $\mathcal{O}(1)\,$Gyr, which may differ from the current values. We therefore conservatively assume a Gaussian systematic uncertainty of $100\,\%$ on the normalization of each neutrino component independently\footnote{In principle, this allows negative normalization for the background, but in practice the data is constraining enough to exclude this situation. The result is that the normalizations of the neutrino fluxes are effectively free.}, including each of the components from different nuclear processes in the Sun.

\paragraph{Backgrounds from $\alpha$-decays  ---}

One possible background is from the `uranium series' of uranium-238, a decay chain which proceeds via a series of $\alpha$ and $\beta$ decays. With each $\alpha$-decay in the series, the child nucleus recoils against the $\alpha$ particle with $\mathcal{O}(10 - 100)\,\mathrm{keV}$ energy. The half-life of $^{238}\mathrm{U}$ is $T_{1/2}\sim4.5\times 10^{9}\,\mathrm{yr}$, while the subsequent decays occur much more quickly ($T_{1/2} \lesssim 2.5 \times 10^5 \,\mathrm{yr}$). Thus, the vast majority of $^{238}\mathrm{U}$ nuclei which decay will have completed the entire decay chain over the age of the mineral (see Ref.~\cite{Baum:2018zzz} for a more in-depth discussion). Even if the $\alpha$ tracks are not observable, the numerous decays in the chain will lead to a characteristic pattern of tracks. We assume that all such track arrangements can be rejected as background. However, we note that in a $10\,\mathrm{mg}$ sample of sinjarite there would be $\mathcal{O}(10^7)$ completed decay chains and further work is required to estimate whether such large rejection factors will be achievable in a real experiment. 

A more problematic background comes from uranium-238 nuclei which have only undergone a single $\alpha$ decay (${^{238}\mathrm{U}} \rightarrow {^{234} \mathrm{Th}} + \alpha$). In this case, the thorium-234 child has a characteristic recoil energy of 72 keV and (assuming that the $\alpha$ track is not seen) is indistinguishable from a DM-induced recoil. The number of such $1 \alpha$-thorium tracks depends on the relative half-lives of the $^{238}\mathrm{U}$ and $^{234}\mathrm{U}$ decays [($^{234}{\rm U} \to {^{230}{\rm Th}} + \alpha$) is the second $\alpha$-decay in the uranium-238 decay chain] and is roughly
\begin{equation}
   n_\mathrm{Th} \approx 10 ^ { 9 }\, \mathrm { kg } ^ { - 1 } \left( \frac { C ^ { 238 } } { 0.01\, \mathrm { ppb } } \right)\,,
\end{equation}
where $C^{238}$ is the uranium-238 concentration (by weight).

It should be possible to estimate the normalization of radioactive backgrounds to a high precision, for example by measuring the number of full $^{238}\mathrm{U}$ decay chains in the sample. We therefore assume a $1\,\%$ systematic uncertainty on the normalization of the $1 \alpha$-thorium background,
though as we will see in Sec.~\ref{sec:Sensitivity}, the projected sensitivity of paleo-detectors is limited more by uncertainties in background shapes.

\paragraph{Neutron-Induced Backgrounds ---}

Fast neutrons produced in or around the target minerals will scatter elastically with nuclei. Such neutrons have a mean free path of a few centimetres and typically give rise to $10-100$ nuclear recoil events with energies comparable to those caused by DM. This large number of neutron-induced tracks is therefore difficult to reject as background.

Fast neutrons may be produced in the spontaneous fission (SF) of $^{238}\mathrm{U}$. This accounts for roughly 1 in every $2 \times 10^6$ decays of $^{238}\mathrm{U}$,  producing $\sim 2$ fast neutrons with MeV energies per SF decay. Neutrons may also be produced in $(\alpha,\,n)$ reactions, in which nuclei absorb an incident $\alpha$ particles and emit fast neutrons. The neutron-induced recoil spectra are estimated using the \texttt{SOURCES-4A} code~\cite{Madland1999}, including both SF and $(\alpha,\,n)$ contributions, as described in Ref.~\cite{Baum:2018zzz}.
As in the case of $1 \alpha$-thorium backgrounds, the normalization of the neutron-induced background scales with uranium-238 contamination. For minerals found in ultra-basic rocks (nchwaningite and olivine), we assume a uranium-238 contamination of $C^\mathrm{238} = 0.1 \,\mathrm{ppb}$ by weight, while for those found in marine evaporites (halite and sinjarite) we assume $C^\mathrm{238} = 0.01 \,\mathrm{ppb}$.

We assume a $1\,\%$ systematic uncertainty  on the normalization of the neutron-induced backgrounds. We discuss the sensitivity of our results to the assumed background uncertainty and explore more extended shape systematics in Sec.~\ref{sec:Sensitivity}.

\subsection{Analysis Theory}
\label{sec:AnalysisTheory}

To estimate projected upper limits and discovery reaches for paleo-detectors, we use the statistical techniques developed in Ref.~\cite{Edwards:2017mnf}. These extend traditional Fisher forecasting methods to the Poisson regime, which allows one to approximate the median result obtained via a Monte-Carlo simulation with minimal computational overhead. We briefly summarize the technique below.

Traditional Fisher forecasting methods are sufficient to accurately calculate expected exclusion limits as well as the discovery reach when in the Gaussian regime. Typically, in direct detection experiments, however, the small number of signal and background events means that the number of expected counts is not well described by a Gaussian distribution. To remain accurate in this Poissonian regime, we adopt the \textit{equivalent counts method}, as developed in Ref.~\cite{Edwards:2017mnf}. The basic procedure for the equivalent counts method is to define a mapping between the expected background counts, their associated uncertainty, and the expected signal such that the full profile log-likelihood is approximated by the Poisson log-likelihood ratio. This mapping is given by,
\begin{align}
  -2\ln \frac{\mathcal{L}_\text{p}(\mathcal{D}_\mathcal{A}(\vect S_0)|\vect S_0)}{\mathcal{L}_\text{p}(\mathcal{D}_\mathcal{A}(\vect S_0)|\vect S)}
  \simeq
  -2\ln \frac{P(b_\text{eq}|b_\text{eq})}{P(b_\text{eq}|s_\text{eq}+b_\text{eq})}
  \;,
  \label{eqn:PoissonApprox}
\end{align}
where $\mathcal{D}_\mathcal{A}(\vect S_0)$ is the Asimov data  set \cite{Cowan:2010js} given no expected signal events and $\vect S$ is the expected number of signal events. $P(a|a')$ represents the Poisson probability mass function, \textit{i.e.}, the probability of $a$ events given $a'$ expected events. The \textit{equivalent} signal and background events are denoted by $s_\text{eq}$ and $b_\text{eq}$, respectively.  They are defined such that the log-likelihood ratio of a simple one-bin Poisson process approximates the full log-likelihood ratio.  
We found expressions for $s_\text{eq}$ and $b_\text{eq}$ in terms of the Fisher matrix of the full problem, which are given in Eq.~(6) and Eq.~(7) of Ref.~\cite{Edwards:2017mnf}.  The procedure leads -- per definition -- to exact results in the limits where the full problem is a one-bin Poisson process or Gaussian, and approximates very well Monte Carlo results in the intermediate range.

The discovery reach and exclusion limit can then, trivially, be calculated by solving
\begin{equation}
\label{eqn:disreach}
  -2\ln \frac{P(s_\text{eq}+b_\text{eq}|b_\text{eq})}{P(s_\text{eq}+b_\text{eq}|s_\text{eq}+b_\text{eq})} = Z^2\;,
\end{equation}
and
\begin{equation}
  s_\text{eq} = Z\sqrt{s_\text{eq} + b_\text{eq}}\;,
  \label{eqn:ULsb}
\end{equation}
respectively, and mapping this back on to the signal parameters of the full model.  The significance level $\alpha$ determines the critical value $Z$, which is derived from the inverse of the standard normal cumulative distribution, denoted $F_{\mathcal{N}}$, as
 \begin{equation}
   Z(\alpha) \equiv F_{\mathcal{N}}^{-1}(1-\alpha)\;.
   \label{eqn:Z}
 \end{equation}
For example, $Z(0.05) = 1.64$ represents a 95\% confidence level. Reference~\cite{Edwards:2017kqw} showed that for most cases the equivalent counts method is accurate to the percent level. In some exceptional cases the derived upper limit exhibits a maximum deviation of 40\% from a fully coverage corrected Monte Carlo calculation, for more detailed discussion see Ref.~\cite{Edwards:2017kqw}. These deviations are most relevant when two distinct regions of the spectrum are present; one in which the signal dominates and the background is in the Poissonian regime, and the other in which the background dominates over the signal. In almost all our cases we expect a significant background so this is unlikely to be a problem.

The \textit{information flux}~\cite{Edwards:2017mnf} provides an intuitive illustration of which region of signal space provide most information about the DM-induced signal. It is a generalized signal-to-noise ratio which allows for the inclusion of extended systematics and covariances. Technically, the information flux is obtained by taking functional derivatives of the Fisher information matrix with respect to the exposure at different track lengths.  It can be thought of as the rate at which the error bar on the parameter of interest will be reduced from an infinitesimal increase in exposure (for a particular track length). We stress that the information flux is used for illustration only and does not enter directly into the calculation of projected upper limits or the discovery reach.

The statistical techniques outlined above are implemented in the software package \texttt{swordfish}\footnote{https://github.com/cweniger/swordfish}~\cite{Edwards:2017kqw}. This package provides a straightforward interface, allowing the user to input signal and background spectra, and efficiently computes the projected exclusion limits and discovery reach. Although the forecasting techniques developed in Ref.~\cite{Edwards:2017mnf} are applicable to unbinned data, the implementation in \texttt{swordfish} requires, for practical reasons, that the data be binned. We use 70 log-spaced bins throughout this work, with the range defined by the resolution as $\sigma_{x_T}/2 \lesssim x_T \lesssim 1000\mathrm{\,nm}$. The number of bins was chosen by incrementally increasing the bin width until the projected constraints on the cross section begin to weaken.  This way we use the minimum number of bins required to resolve all features in the track length spectra. We do this both to minimize computation time and to ensure that the systematics study in Sec.~\ref{sec:Sensitivity} is conservative (using a small number of bins enhances the impact of bin-to-bin variations in the backgrounds).  The same number of bins is used in both the high resolution and high exposure cases to allow for an easier comparison between the two read-out methods.

\subsection{Track Length Spectra}
\label{sec:Spectra}

\begin{figure*}[th!]
    \centering
    \includegraphics[width=0.48\linewidth]{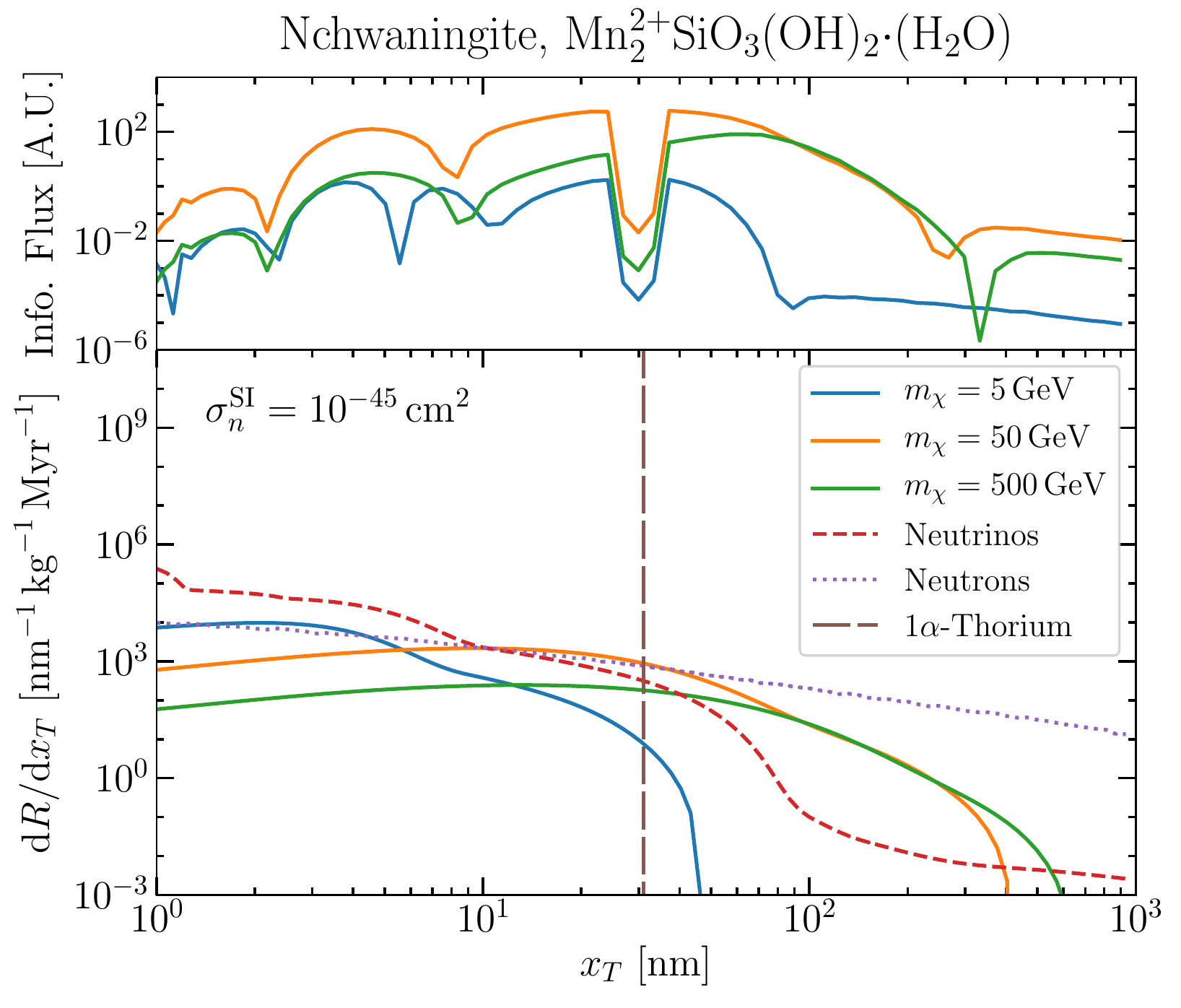}
    \includegraphics[width=0.48\linewidth]{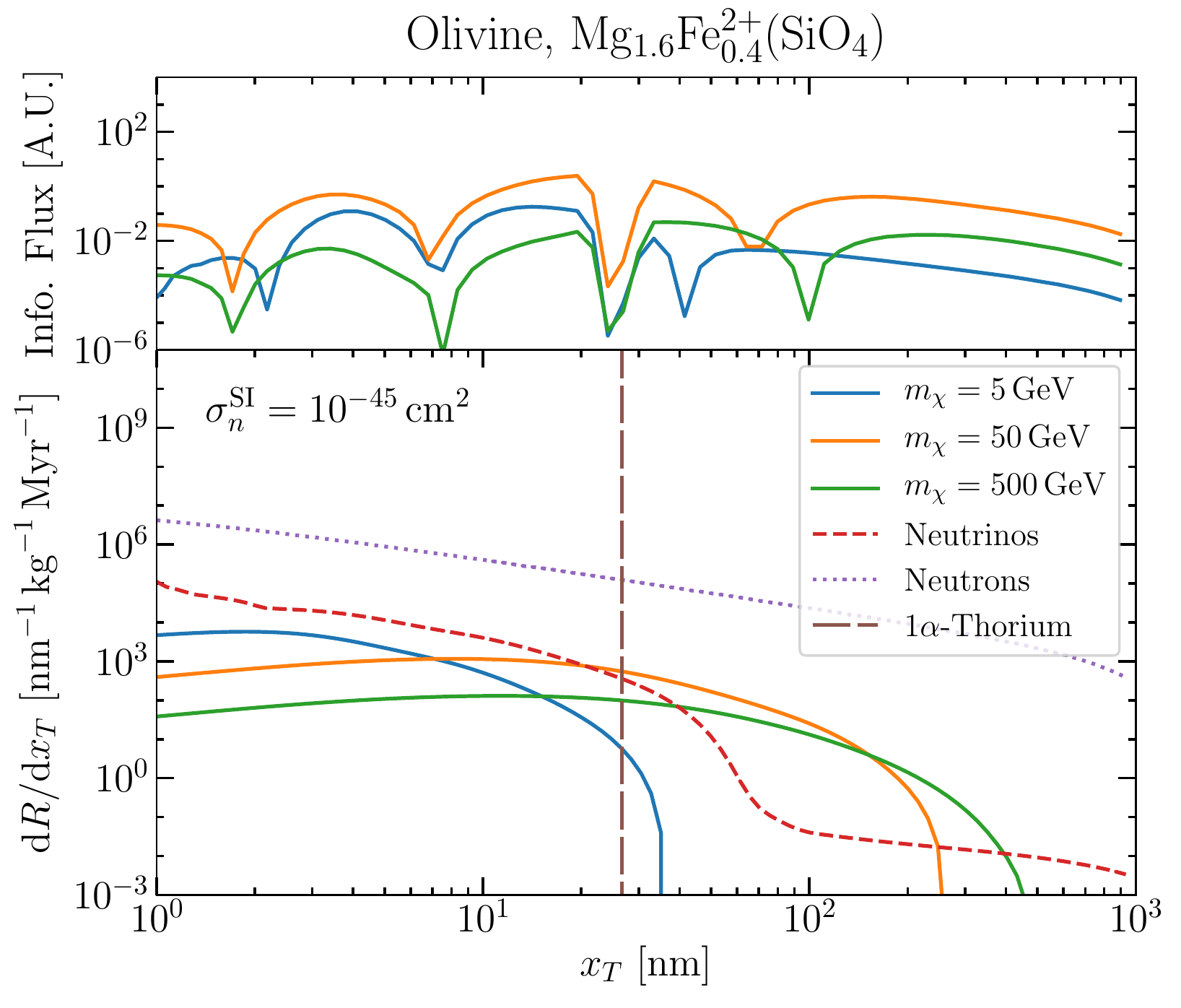}
    \includegraphics[width=0.48\linewidth]{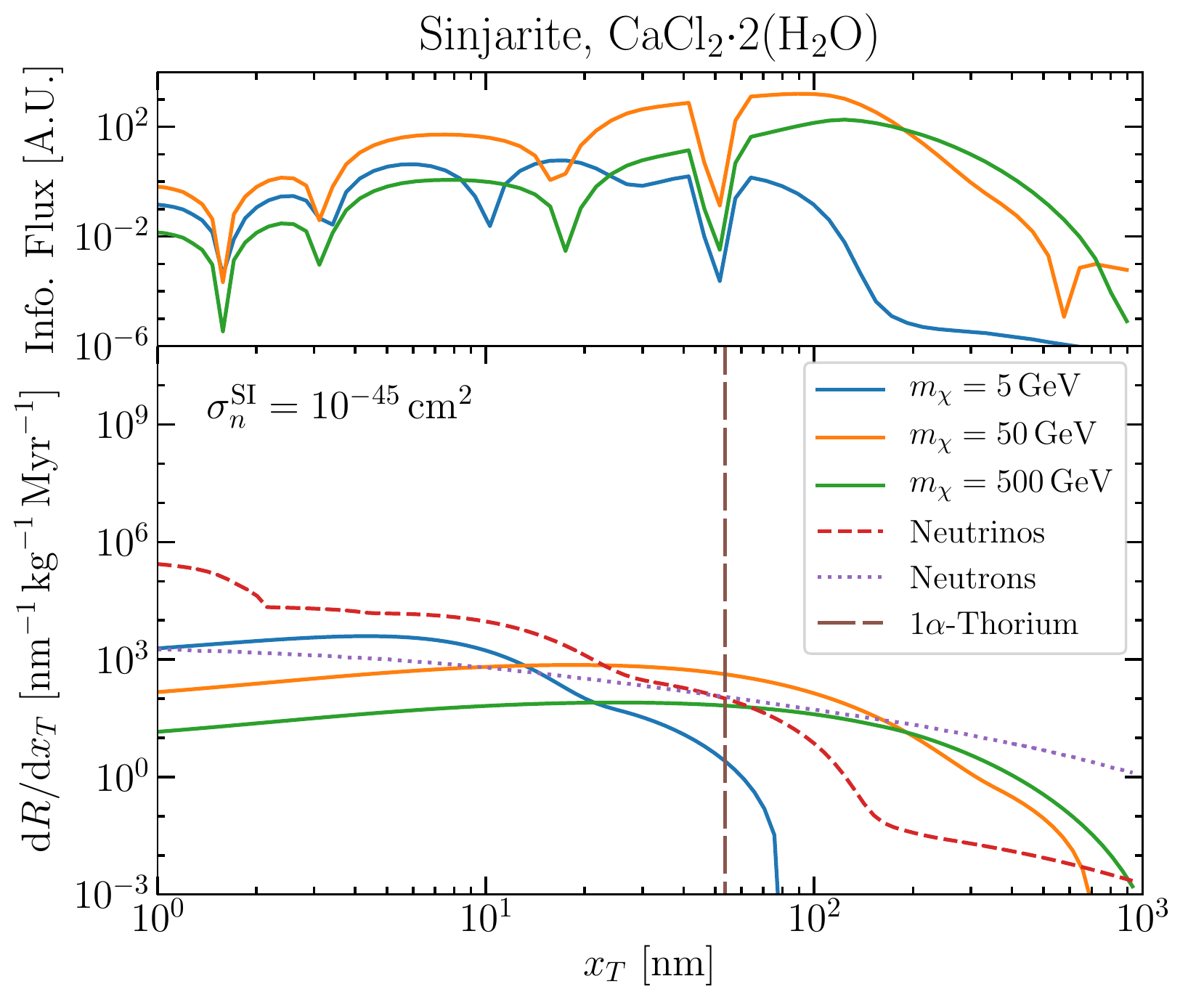}
    \includegraphics[width=0.48\linewidth]{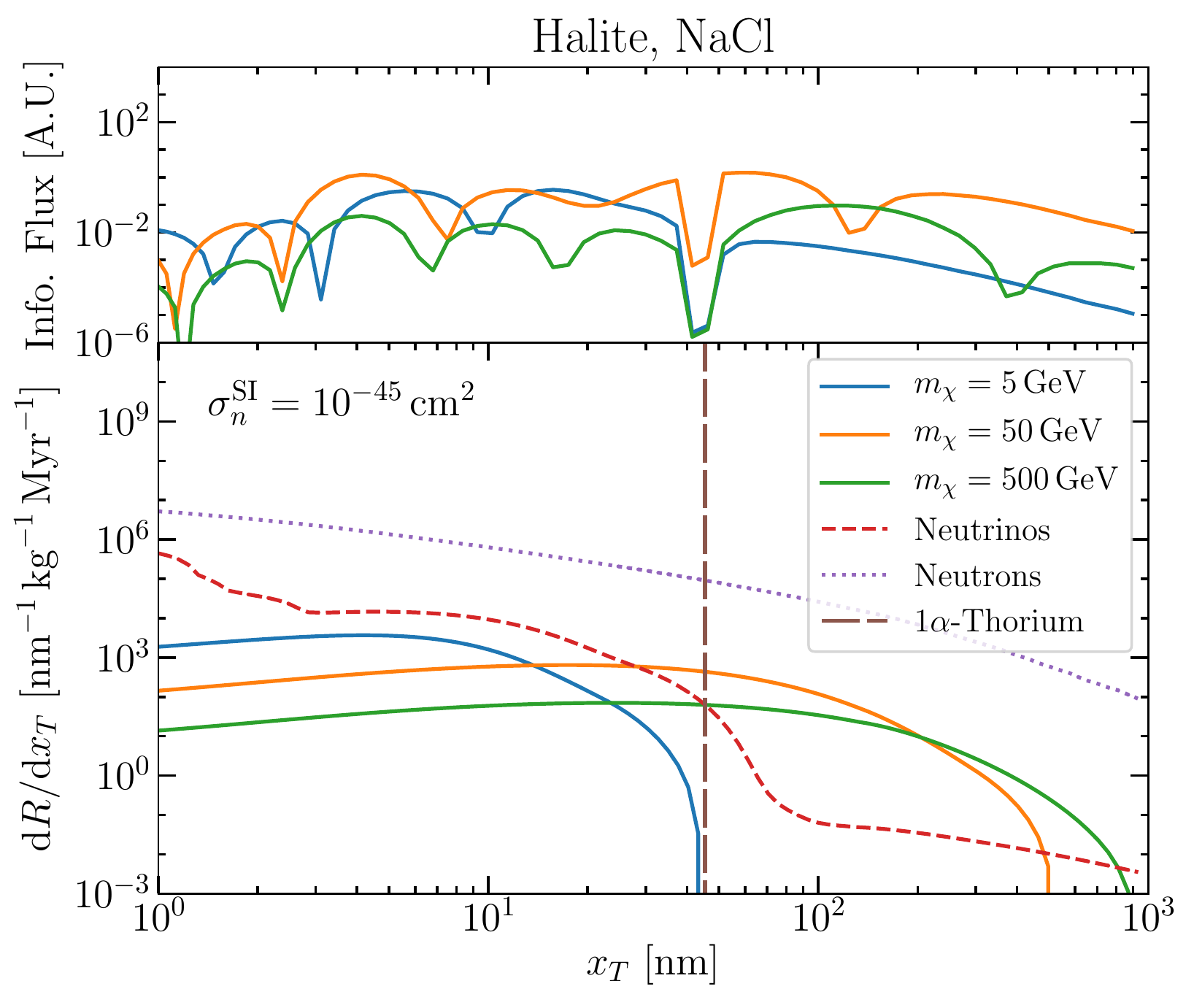}
    \caption{\textbf{Track-length spectra and information flux for three different WIMP masses.} Each panel shows the spectrum of track lengths $x_T$ expected for a different mineral target. WIMP signals are shown as solid lines, while background distributions are shown as dotted and dashed lines. We fix the signal normalization to $\sigma_n^\mathrm{SI} = 10^{-45}\,\mathrm{cm}^2$. The information flux, shown above each set of spectra, is a generalized signal-to-noise ratio discussed in more detail in Sec.~\ref{sec:AnalysisTheory}. The information flux has been calculated for the high-resolution case, with a track-length resolution of $\sigma_{x_T} = 1\,\mathrm{nm}$.}
    \label{fig:SpectraAndIF}
\end{figure*}

We now present the distribution of track lengths expected for DM signals as well as the backgrounds already described in this section. These are shown in Fig.~\ref{fig:SpectraAndIF}, with each panel showing a different mineral. We fix the DM--nucleon cross section to $\sigma_n^\mathrm{SI} = 10^{-45}\,\mathrm{cm}^2$ and show the DM signal spectra for three different DM masses, $m_\chi = 5,\,50\text{ and } 500 \,\,\mathrm{GeV}$, as solid lines. We see that at short track lengths, $x_T \lesssim 10\,\mathrm{nm}$, where signals of lighter WIMPs would appear, the dominant background often comes from solar neutrinos. At longer track lengths, where the signal from heavier DM typically peaks, the dominant backgrounds are radioactive: mono-energetic $1 \alpha$-thorium recoils  and neutron-induced recoils. 

In the upper part of each panel, we also plot the information flux, as described in Sec.~\ref{sec:AnalysisTheory}. The most pronounced feature is a sharp drop for track lengths corresponding to $1 \alpha$-thorium recoils. Little information about a DM signal can be gained by studying tracks of this length, as signal events are here degenerate with $1 \alpha$-thorium tracks.  The various peaks in the information flux indicate the track lengths that provide the most constraining power for the DM signal. These maxima appear either when the signal is large, corresponding to a large signal-to-background ratio, or when one of the backgrounds is prominent. In the latter case, track lengths corresponding to peaks in the information flux allow us to constrain the normalization of a particular background component. This in turn leads to improved constraints on the signal. In between these peaked regions, the information flux is typically suppressed.  Note that the detailed shape of the information flux depends significantly on the specific assumptions that are made about the correlation of background systematics.

We discuss the track length spectra, information fluxes and their impact on paleo-detector sensitivity in more detail in the next section.

\section{Projected Sensitivity}
\label{sec:Sensitivity}

Here we present the results of the sensitivity analysis. First we discuss a simple benchmark case in which we consider systematic errors only on the normalization of the individual background components. We also use this scenario for the mass reconstruction projections in Sec.~\ref{sec:Reconstruction}. In addition, we also consider bin-to-bin systematics in order to assess the impact of shape uncertainties of the background spectra on the projected sensitivity. 

\subsection{Background Normalization Systematics}
\label{sec:Optimistic}

The backgrounds described in Sec.~\ref{sec:Theory} all have an associated uncertainty which must be accounted for within the analysis framework. For our background normalization systematics scenario we assume that the shapes of the signal and backgrounds are fixed with only a systematic uncertainty on the normalization of each background component. The systematic uncertainties we assign to the normalization of different backgrounds are detailed in Sec.~\ref{sec:Backgrounds}. We ignore covariances between the signal and background and between individual backgrounds components. With careful calibration, we may be able to understand the shape of the background to a high degree of precision. In practice, it should be straightforward to produce target samples with high levels of radioactivity-induced tracks in the laboratory, though such an approach is more challenging for the neutrino-induced backgrounds.

Uncertainties in the normalization of the backgrounds can be mitigated when a good `control region' is available, where the signal is sub-dominant and the overall background rate can be well-constrained. This is typically the case for the broad distribution of neutron-induced recoils; even for heavy WIMPs, the signal drops off well below $1000\,\mathrm{nm}$, providing a good control region at large track lengths. Instead, neutrino-induced backgrounds may mimic the DM signal for certain DM masses. If this is the case, no control region is available and limits are severely weakened.

\begin{figure*}[th!]
    \centering
    \includegraphics[width=0.49\linewidth]{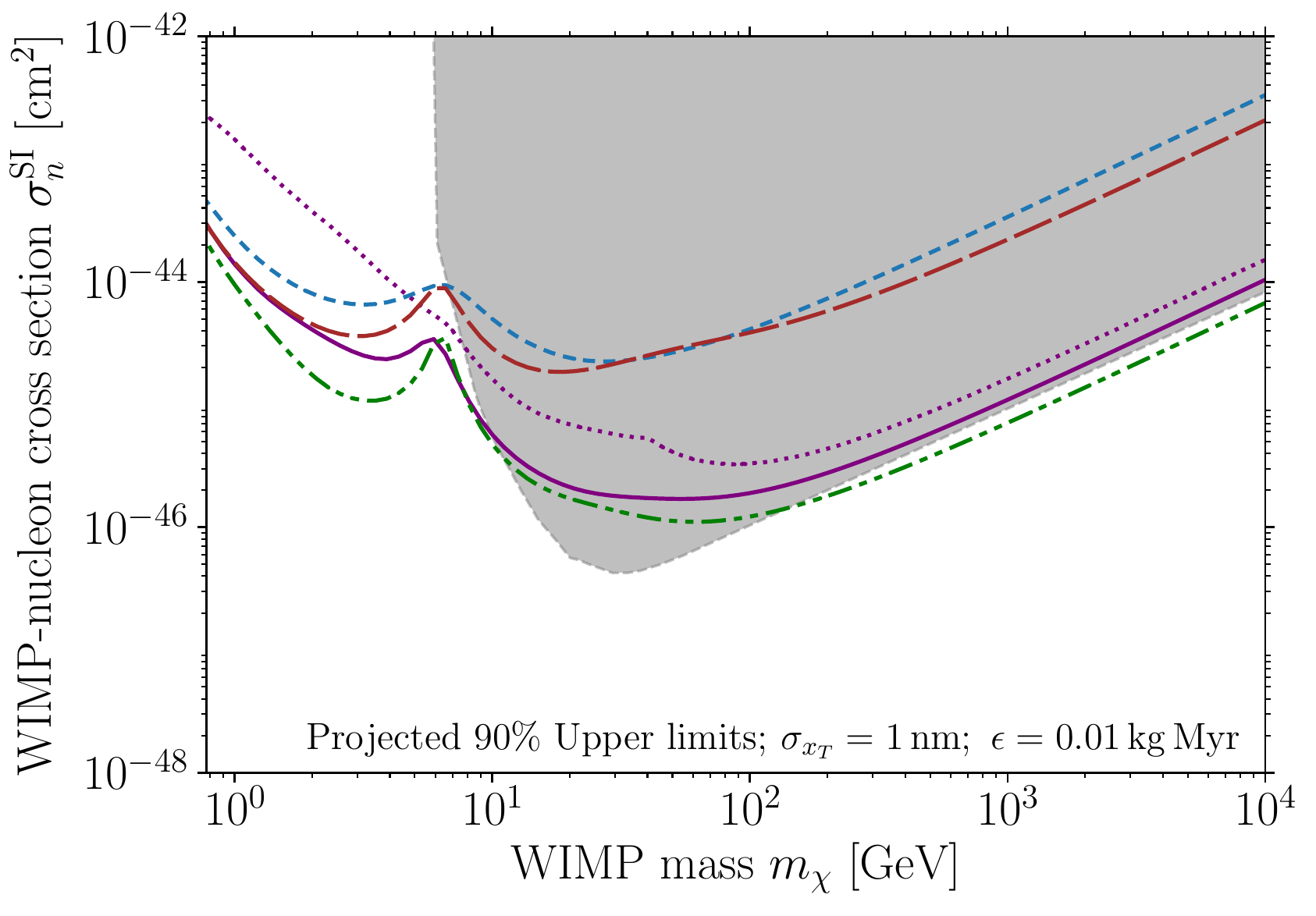}
    \includegraphics[width=0.49\linewidth]{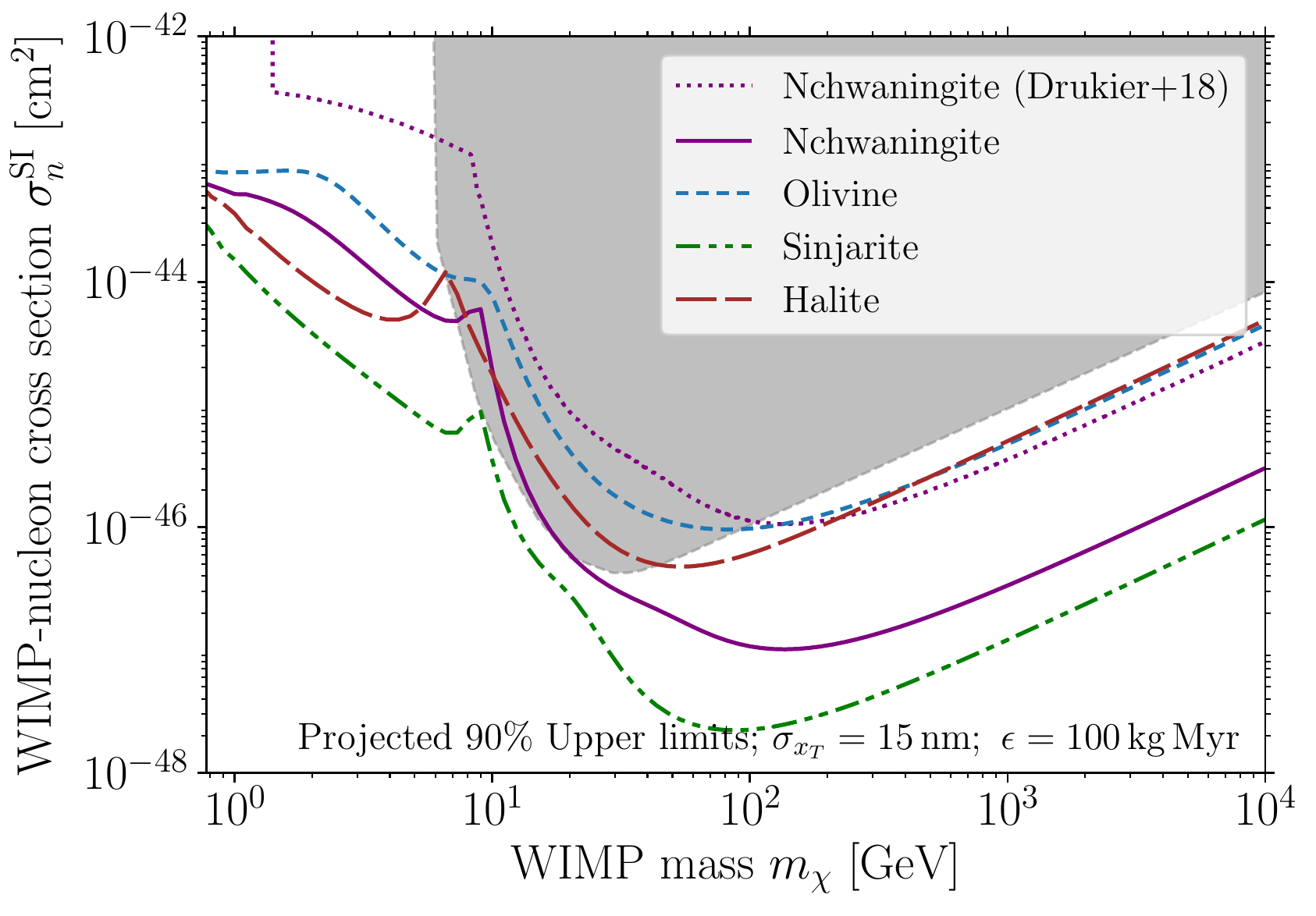}
    \includegraphics[width=0.49\linewidth]{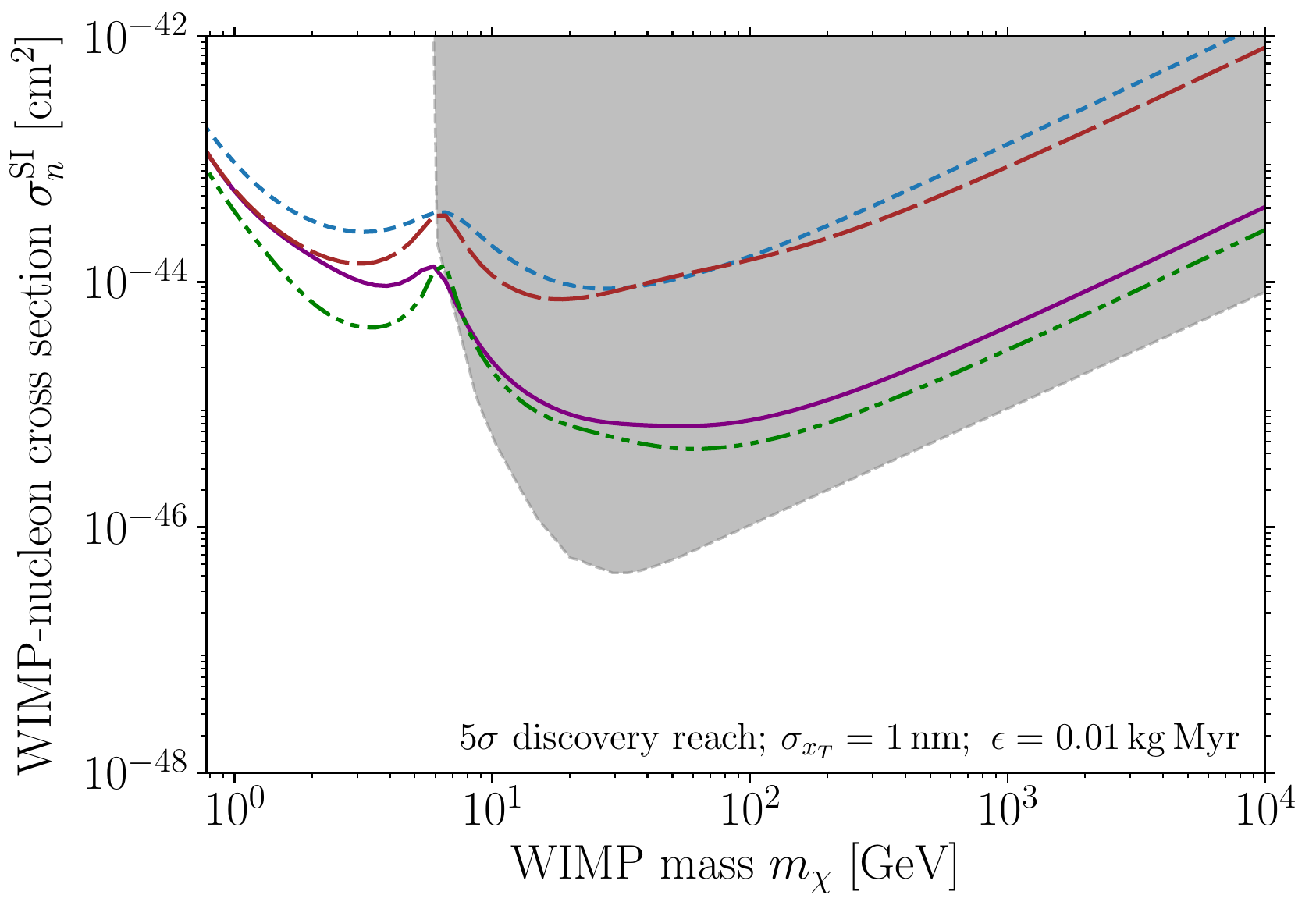}
    \includegraphics[width=0.49\linewidth]{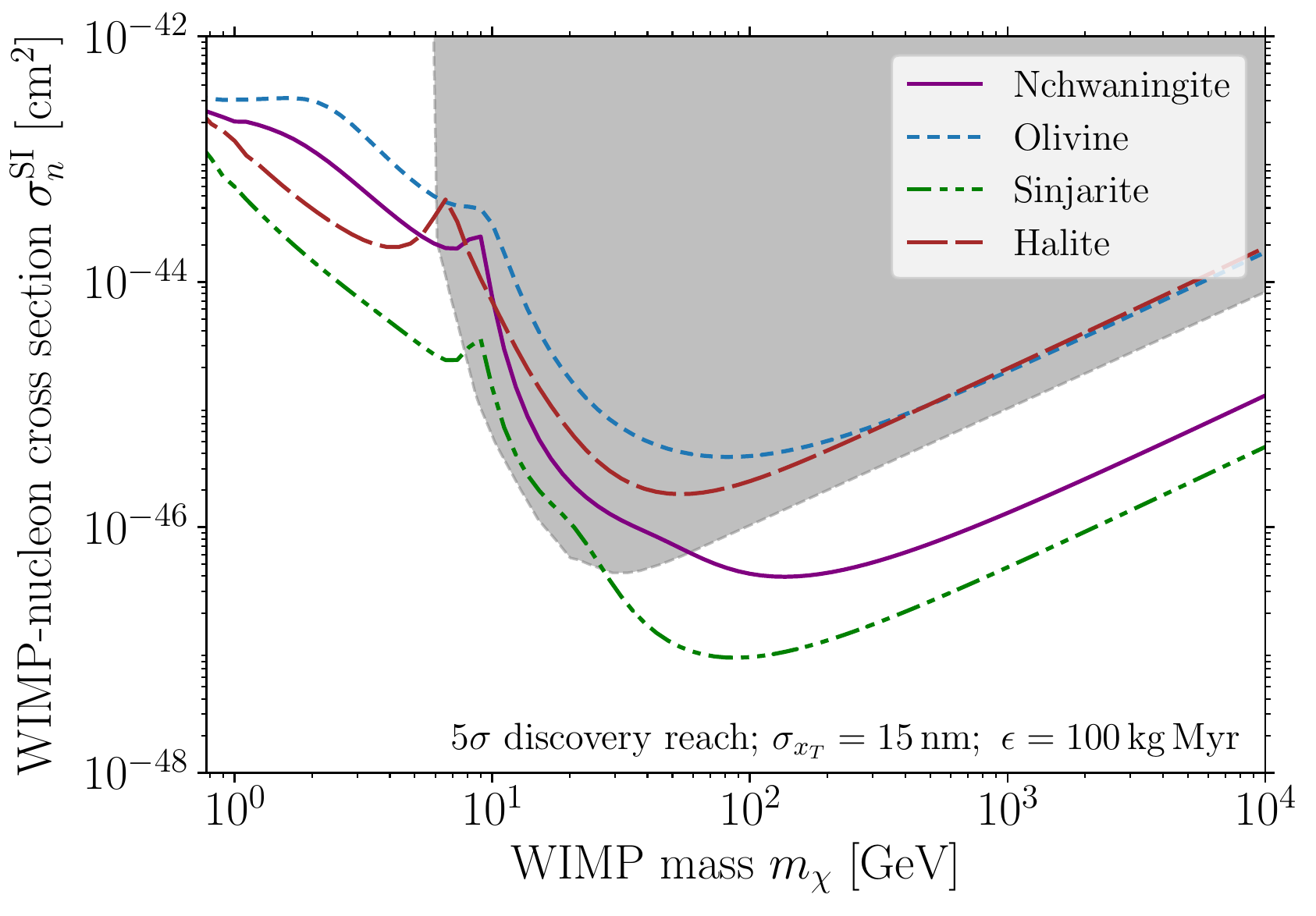}
    \caption{\textbf{Projected paleo-detector sensitivity using a spectral analysis. Top:} Projected 90\% Upper Limits. \textbf{Bottom:} $5\,\sigma$ Discovery Reach. In the \textit{left column}, we assume $\sigma_{x_T} = 1\,\mathrm{nm}; \,\,\epsilon = 0.01\,\mathrm{kg}\,\mathrm{Myr}$ (high resolution case). In the \textit{right column}, we assume $\sigma_{x_T} = 15\,\mathrm{nm}; \,\,\epsilon = 100 \,\mathrm{kg}\,\mathrm{Myr}$ (high exposure case). We assume a 100\,\% systematic uncertainty on the normalization of each individual neutrino component, as well as a 1\,\% systematic uncertainty on the normalization of both the neutron background and the $1 \alpha$-thorium background. \textit{Gray regions} show the parameter space currently excluded by conventional direct detection experiments~\cite{Agnese:2018gze,Aprile:2018dbl}. In the \textit{upper panels}, we also show the projected limits from nchwaningite using a sliding window analysis (dotted purple), as reported recently in Ref.~\cite{Baum:2018zzz}.}
    \label{fig:limits}
\end{figure*}

We show in Fig.~\ref{fig:limits} the projected sensitivity for the four minerals we consider in this work. The top two panels show the projected 90\% confidence exclusion limits while the bottom two panels show the projected $5\,\sigma$ discovery reach~\cite{Billard:2011zj}, which we define as the line above which the paleo-detector setups we consider would have at least 50\,\% chance of achieving a $5\,\sigma$ discovery of DM. In gray, we show current bounds from conventional direct detection experiments, coming predominantly from XENON1T~\cite{Aprile:2018dbl} and SuperCDMS~\cite{Agnese:2017jvy} in this mass and cross section range.

In the left panels, we show the high resolution case, with $\sigma_{x_T} = 1\,\mathrm{nm}$ and an exposure of $\epsilon = 0.01 \,\mathrm{kg} \, \mathrm{Myr}$. The `bump' in the limits at $m_\chi \sim 7\,\mathrm{GeV}$ is due to the WIMP signal spectrum near this mass being mimicked by the spectra of $^8\mathrm{B}$  and hep neutrinos.
This `Solar neutrino floor' has been studied in detail in, for example, Refs.~\cite{Billard:2013qya,Strigari:2016ztv,OHare:2016pjy}. The loss in sensitivity is even more pronounced in our case, owing to the larger systematic uncertainty we have assigned to the Solar neutrino flux. In addition, moving away from $m_\chi \sim 7\,\mathrm{GeV}$, the spectra are no longer degenerate, meaning that control regions are rapidly regained and sensitivity restored. In contrast to conventional direct detection experiments, paleo-detectors have large enough exposures to directly constrain the normalization of the $^8$B/hep Solar neutrino fluxes.

Even accounting for these low energy background neutrinos, in the left panels, the limits at low mass are substantially stronger than limits from any conventional detector. This is due to the ability of modern imaging techniques to measure small track lengths. Tracks of $1\,\mathrm{nm}$ in length typically correspond to recoil energies around 100 eV, giving a threshold comparable to the CRESST-2017 Surface Run~\cite{Angloher:2017sxg}, albeit with a much larger exposure. Thus, for DM lighter than $\sim 10\,$GeV, paleo-detectors may probe DM--nucleon cross sections much smaller than the projected sensitivity of conventional direct detection experiments.

For larger DM masses, the sensitivity is severely limited by neutron-induced backgrounds. Nchwaningite and sinjarite contain hydrogen, which is an efficient moderator of fast neutrons. Thus, the rate of neutron-induced backgrounds is much smaller than in olivine and halite, which do not contain hydrogen. In addition, for halite and sinjarite we assume an intrinsic contamination of $C^{238} = 0.01 \,\mathrm{ppb}$, whereas for olivine and nchwaningite we assume $C^{238} = 0.1\,$ppb. For nchwaningite and sinjarite, the minerals containing hydrogen, cross sections as low as $10^{-46}\,\mathrm{cm}^2$ could be probed for a DM mass of 50\,GeV, assuming an exposure of $0.01 \,\mathrm{kg} \, \mathrm{Myr}$, as shown in the left panels. At higher masses, the projected sensitivity of paleo-detectors is comparable to current XENON1T bounds.

In the right panels of Fig.~\ref{fig:limits}, we show the projected sensitivity for the lower resolution case, with $\sigma_{x_T} = 15\,\mathrm{nm}$ and a larger exposure of $\epsilon = 100 \,\mathrm{kg} \, \mathrm{Myr}$. For DM heavier than $100 \, \mathrm{GeV}$, it is possible to probe DM--nucleon cross sections a factor of 30 and 100 smaller than current XENON1T bounds using nchwaningite and sinjarite, respectively. For DM lighter than $10 \, \mathrm{GeV}$, the sensitivity is marginally weaker than in the high resolution case; at low DM mass, the signal spectrum peaks towards shorter track lengths and is thus challenging to resolve with worse resolution.

We note that the projected limits are comparatively weak between 20 and 100 GeV. This is because the peak due to $1 \alpha$-thorium recoils (vertical dashed lines in Fig.~\ref{fig:SpectraAndIF}), typically coincides with the peak in the signal spectra in this mass range\footnote{For $m_\chi \sim 50\,\mathrm{GeV}$, the typical recoil energy for nuclei in the minerals we consider is $\sim 20 \,\mathrm{keV}$, much smaller than the $72\,\mathrm{keV}$ $1 \alpha$-thorium recoil. However, the stopping power for lighter nuclei such as $\mathrm{Ca}$, $\mathrm{Cl}$ and $\mathrm{Na}$ is smaller, leading to similar track lengths.}. We see in the upper panels of Fig.~\ref{fig:SpectraAndIF} that the broad peak in the information flux for the 50 GeV case typically occurs at the same position as the `dip' caused by the $1 \alpha$-thorium background.

We now compare our results with those obtained using a simpler `sliding window' analysis in Refs.~\cite{Baum:2018tfw,Baum:2018zzz}, where a cut-and-count analysis is performed over a window chosen to optimize the signal-to-noise ratio. In the top two panels of Fig.~\ref{fig:limits}, we show the limit obtained in Ref.~\cite{Baum:2018zzz} for nchwaningite.

In the high resolution case (left panels of Fig.~\ref{fig:limits}), we find that the `sliding window' analysis is only marginally less sensitive at the highest masses. Near $m_\chi \sim 50\,\mathrm{GeV}$, the full spectral analysis is more sensitive by a factor of a few due to better rejection of the sharply peaked $1 \alpha$-thorium background. Going to lower masses, the `bump' corresponding to the $^8$B/hep Solar neutrinos becomes more pronounced in a full spectral analysis. For lighter DM, the neutrino- and DM-induced spectra become distinguishable again. The spectral analysis can effectively reduce the error on the normalization of the neutrino-induced background by using information from track lengths where neutrino-induced events dominate. For example, at masses of $1\,\mathrm{GeV}$ this allows the projected sensitivity from the spectral analysis to improve nearly an order of magnitude with respect to that found in the `sliding window' analysis. 

In the high exposure case (right panels of Fig.~\ref{fig:limits}), the spectral analysis gains roughly an order of magnitude in sensitivity with respect to the 'sliding window' analysis for DM heavier than $\sim 100\,$GeV. The lower resolution makes it more difficult to exploit subtle differences in the shape of the signal and background spectra. However, at the longest track lengths considered, the tracks induced by neutrons always dominate over those induced by DM, cf. Fig.~\ref{fig:SpectraAndIF}. Because of the larger exposure, this `control region' has sufficient statistics to tightly constrain the normalization of the neutron-induced background. Thus, the sensitivity to higher mass WIMPs is improved  with respect to the `sliding window' analysis. 

\subsection{Background Shape Systematics}

\begin{figure}[tb!]
    \centering
    \includegraphics[width=\linewidth]{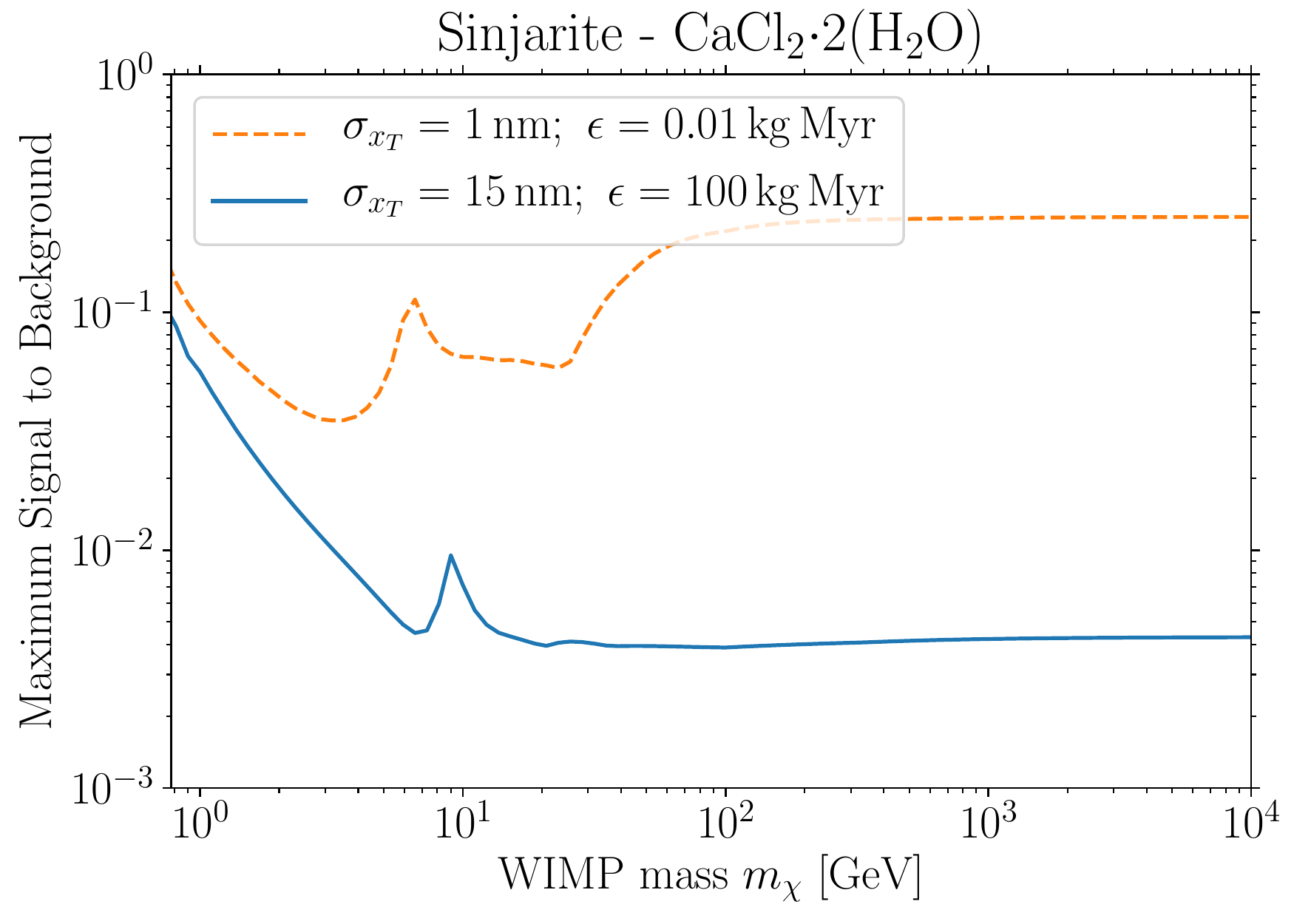}
    \caption{\textbf{Maximum signal-to-background ratio at the 90\,\% confidence limit.} Both lines show the maximum signal-to-background ratio (over all track lengths) for sinjarite with $\sigma_n^{\mathrm{SI}}$ set to the projected 90\,\% confidence exclusion limit at each mass. The \textit{orange (dashed) line} shows the high-resolution case. Here we see the signal-to-background is relatively constant around $10-30\,$\%. The \textit{blue (solid) line} shows the high-exposure \,case. At low masses the maximum signal-to-background is roughly 10\,\% whereas at high masses this is reduced to $0.4-0.5\,$\%. This is reflected by the increased sensitivity of the limit to bin-to-bin background shape uncertainties, as shown in Fig.~\ref{fig:Systematics_check}.}
    \label{fig:StoB}
\end{figure}

As discussed in the previous section, our sensitivity to DM at high masses is typically limited by neutron-induced backgrounds. Conversely, neutrinos are the dominant background for low DM mass. As we show in Fig.~\ref{fig:StoB}, the maximum signal-to-background ratio (as a function of track length) for the high exposure case is typically much smaller than 10\,\%. This means that the strength of the limits depends in principle on a \%-level understanding of the shape of the background distributions. For the high resolution case the maximum signal-to-background is typically closer to 30\,\%. Therefore, the projected sensitivity should be more robust to shape uncertainties in that case.

In order to explore how the sensitivity of paleo-detectors is affected by such background shape uncertainties, we assign a Gaussian systematic error to the normalization of each bin of each background component. Our analysis therefore allows the backgrounds in individual bins to fluctuate independently. Because we no longer assign systematic uncertainties to the overall normalization of the backgrounds, in some situations (e.g. when the bin-to-bin uncertainty is chosen to be smaller than the normalization uncertainty in Sec.~\ref{sec:Optimistic}) the projected limit with shape uncertainties may be stronger than in the normalization systematics case. In such situations, we set the projected limit equal to the normalization systematics case.

\begin{figure}[t!]
    \centering
    \includegraphics[width=\linewidth]{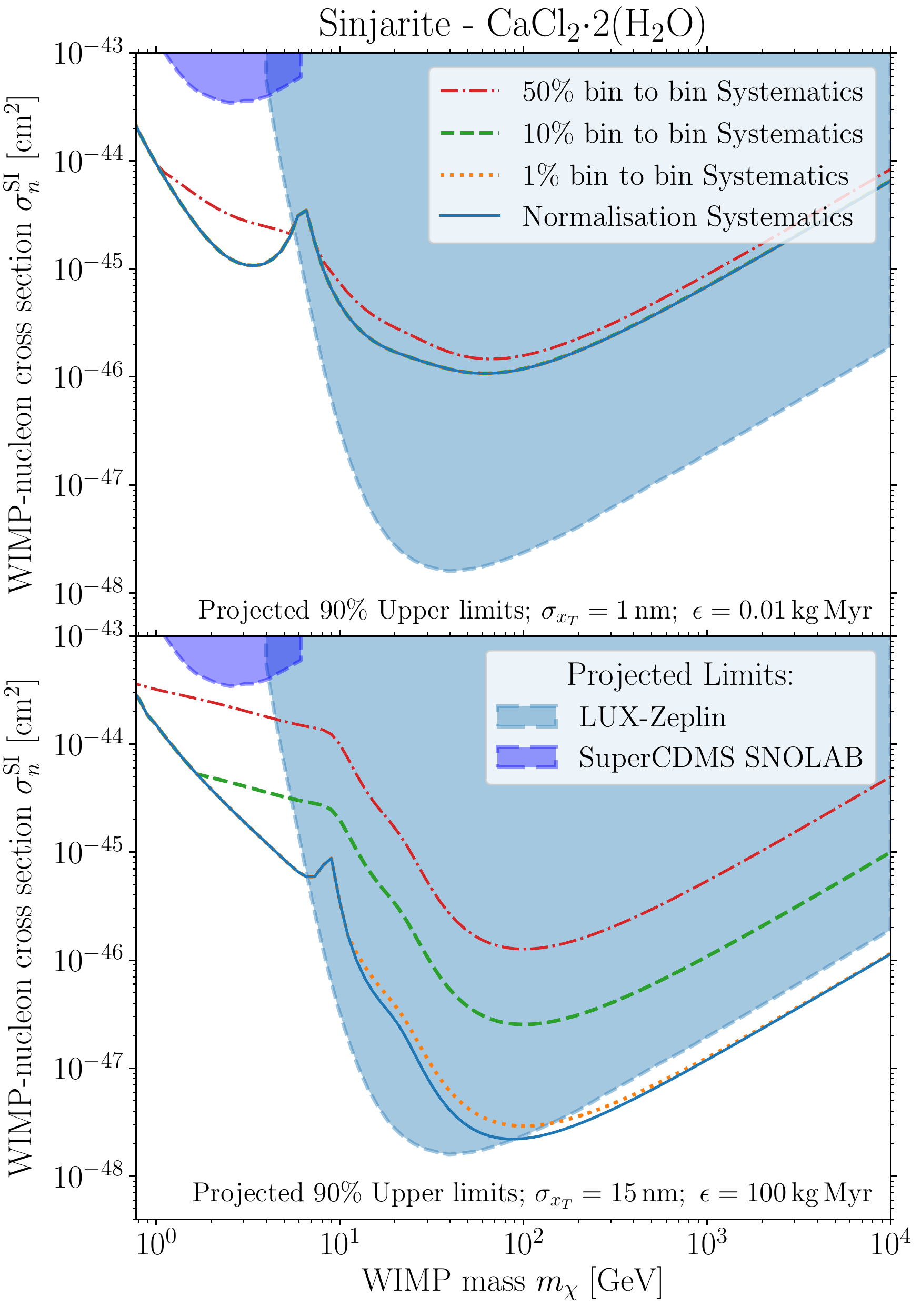}
    \caption{\textbf{Projected 90\,\% confidence limits for sinjarite including background shape systematics.} Top and bottom panels show the high resolution and high exposure cases, respectively.  \textit{Blue line:} Background normalization systematics case with systematic normalization uncertainties for each background component. The normalization systematic on neutrinos here is set to 100\,\% whereas for the radioactive backgrounds we assume 1\,\% normalization error.  \textit{Red, Orange, and green lines:} Background shape systematics case where we allow the normalization of each background component in each of the 70 log-spaced track-length bins to fluctuate independently. The red, green, and orange lines show results for 50\,\%, 10\,\%, and 1\,\% systematic uncertainty per bin, respectively. Note that where the bin-to-bin systematics produce a limit stronger than that of the normalization systematics case, we set the projected limit assuming normalization systematics. In the top panel the 1\% (orange dotted) and 10\,\% (green dashed) bin-to-bin systematic lines are therefore not distinguishable from the normalization systematics case (blue solid). Shaded regions show projected limits from LUX-Zeplin~\cite{Akerib:2018lyp} and SuperCDMS SNOLAB (Ge)~\cite{Agnese:2016cpb}.}
    \label{fig:Systematics_check}
\end{figure}

In Fig.~\ref{fig:Systematics_check}, we plot the limits obtained with various bin-to-bin background systematics for a sinjarite paleo-detector. For comparison, we also show the sensitivity obtained in the background normalization systematics case described in the previous sub-section. These results are mostly illustrative since the actual level of shape uncertainties is hard to anticipate \emph{a priori}.  However, they indicate the level of uncertainty that can be tolerated in practice.

As expected, allowing some variation in the shape of the backgrounds degrades the sensitivity of paleo-detectors. In the upper panel of Fig.~\ref{fig:Systematics_check} we show the high resolution case. The limit is unaffected by bin-to-bin systematics until they are increased to 50\,\%. The more relevant uncertainty is therefore the overall normalization of the background components.

The high exposure case is shown in the lower panel of Fig.~\ref{fig:Systematics_check}. We find a much greater dependence on the background shape systematics. For 10\,\% bin-to-bin systematic uncertainties, the sensitivity is degraded by an order of magnitude compared to the background normalization systematics case for DM heavier than $\sim40\mathrm{\,GeV}$. Another factor of $\sim 5$ is lost when increasing the bin-to-bin systematics from 10\,\% to 50\,\%.

The high resolution case is more robust to background shape systematics primarily because of its larger signal-to-background ratio, as shown in Fig.~\ref{fig:StoB}. In addition, spectral features in the high resolution case are more pronounced, allowing for easier distinction between signal and background even when there are significant uncertainties in the background shapes.

For comparison with our projections, we also show in Fig.~\ref{fig:Systematics_check} the projected 90\,\% confidence exclusion limits from LUX-Zeplin~\cite{Akerib:2018lyp} and SuperCDMS SNOLAB (Ge)~\cite{Agnese:2016cpb} (planned for data-taking from 2020 onwards), with respective exposures of $1.5\times10^4 \mathrm{\,kg\,yr}$ and $44\mathrm{\,kg\,yr}$. For DM masses below $10 \mathrm{\,GeV}$, the high resolution case can improve upon future SuperCDMS SNOLAB constraints by up to one and a half orders of magnitude. For the case of 50\,\% bin-to-bin shape systematics, sinjarite would still be an order of magnitude more sensitive than SuperCDMS SNOLAB projections at $2 \mathrm{\,GeV}$. The high exposure case can achieve the same sensitivity as LUX-Zeplin to higher mass DM only if the background shape uncertainties are kept at the 1\,\% level.

\section{Constraining the Dark Matter Mass}
\label{sec:Reconstruction}

In this section, we investigate to what extent the properties of a DM candidate, in particular its mass, could be constrained in the hypothetical case of a DM discovery. Thus, we switch from projecting limits, as in Sec.~\ref{sec:Sensitivity}, to parameter reconstruction. A priori there is no reason for the DM to appear in any particular region of the parameter space. Instead of employing benchmark scenarios as often done in the literature~\cite{Peter:2011eu,Strege:2012kv,Peter:2013aha,Bozorgnia:2018jep}, we perform a benchmark-free study using the Euclideanized signal method \cite{Edwards:2017kqw,Edwards:2018lsl}.

The Euclideanized signal method maps points in the model parameter space -- here, ($m_\chi$, $\sigma_n^\mathrm{SI}$) -- onto points in a `Euclideanized signal' space, taking into account systematic uncertainties,  covariances and nuisance parameters. Likelihood ratios between points in the model space are mapped to Euclidean distances in the `Euclideanized signal' space. Comparing the Euclidean distances between large numbers of points is computationally fast (using clustering algorithms), allowing us to efficiently map out the reconstruction prospects over a wide range of the parameter space. Full details of the Euclideanized signal method can be found in Refs.~\cite{Edwards:2017kqw,Edwards:2018lsl} and a summary is given in App.~\ref{sec:EuclMeth}.

For regions where a future signal would provide a closed constraint on the DM mass, we calculate the accuracy to which this is possible by defining the fractional uncertainty as 
\begin{equation}
\label{eq:massrecon}
    \frac{\Delta m_{\chi}}{m_{\chi}} = \frac{|m_{\chi,\mathrm{max}}-m_{\chi,\mathrm{min}}|}{m_{\chi}}\;.
\end{equation}
Here, $m_{\chi,\mathrm{max}}$ and $m_{\chi,\mathrm{min}}$ are the maximum and minimum edges of the two-dimensional $2\,\sigma$ confidence contour around a point with a given $(m_{\chi},\sigma_n^\mathrm{SI})$. Note that these contours are typically quite asymmetric, usually extending much further towards masses larger than the true mass than towards masses smaller. Thus, a fractional uncertainty $\Delta m_\chi/m_\chi \gtrsim 1$ does not necessarily imply that no information about the DM mass can be obtained. Rather, $\Delta m_\chi/m_\chi \gtrsim 1$ typically implies that $\Delta m_\chi/m_\chi \sim m_{\chi, \max}/m_\chi$, while usually $m_{\chi, \min}$ is not much smaller than $m_\chi$.

\begin{figure}[t!]
    \centering
    \includegraphics[width=\linewidth]{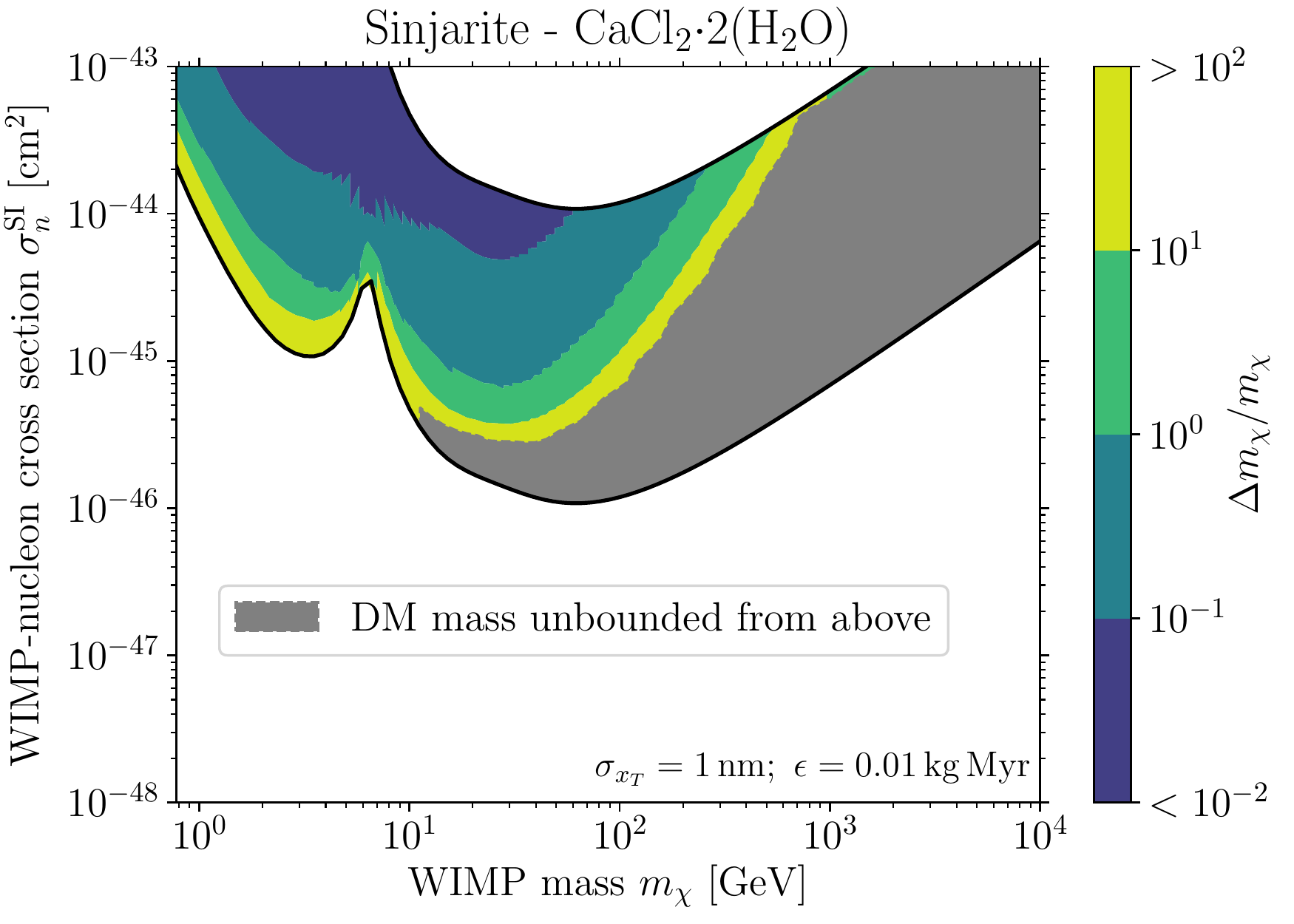}
    \includegraphics[width=\linewidth]{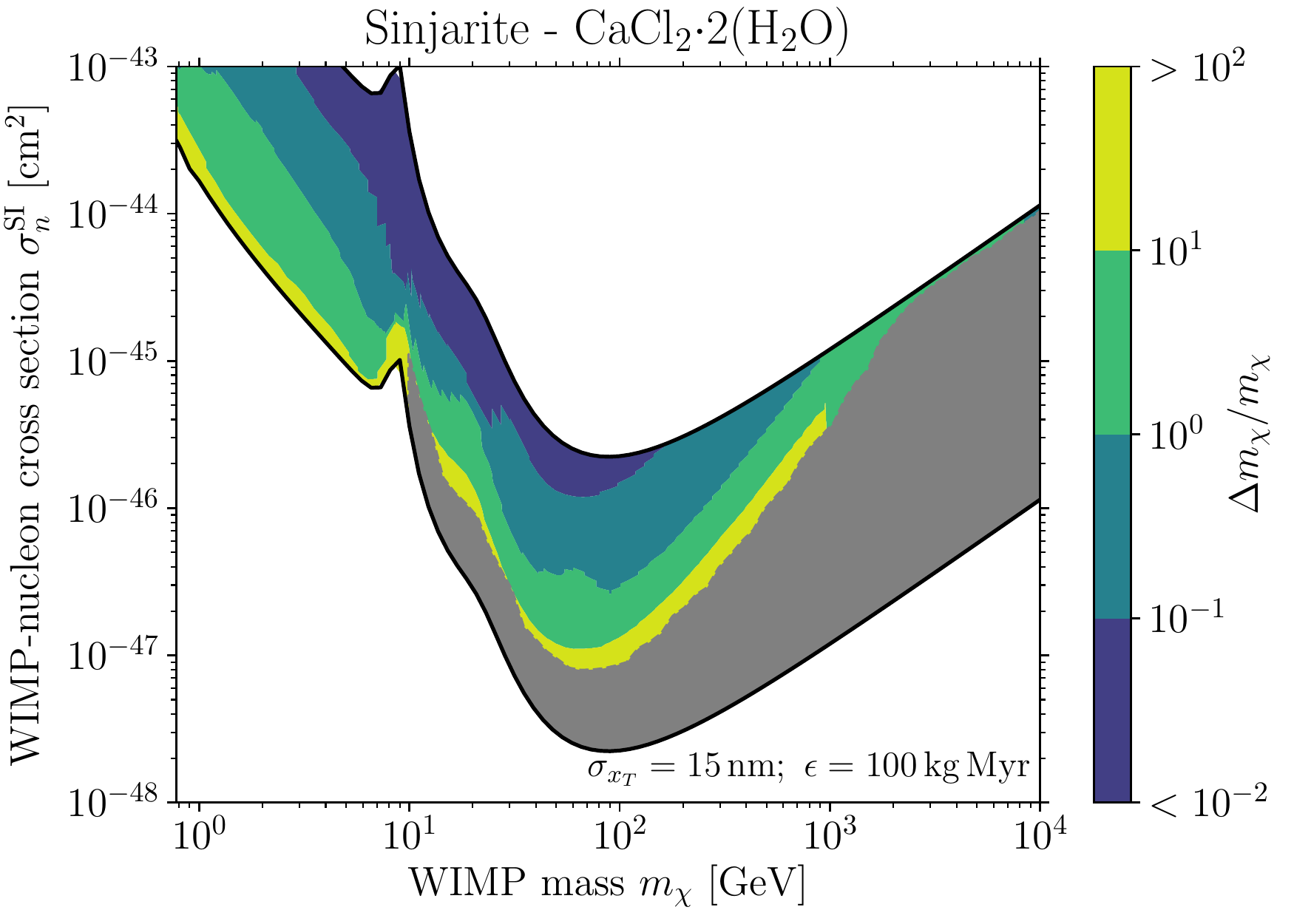}
    \caption{\textbf{Constraints on the mass of a DM particle from a future signal.} Gray shaded regions correspond to parameter points where the DM mass is unconstrained from above at the $2\,\sigma$-level. The colored contours indicate the fractional uncertainty on the DM mass obtained by constraining a future signal, as defined in Eq.~\eqref{eq:massrecon}. Top and bottom panels show the high resolution and high exposure cases, respectively. The lower black lines in both panels correspond to the projected 90\,\% confidence limit in Fig.~\ref{fig:limits}. We consider regions between these lower lines and a factor of 100 larger, indicated by the upper black lines. Note that some of these regions are already excluded by current experiments (see shaded regions in Fig.~\ref{fig:limits}).}
    \label{fig:mass_reconstruction}
\end{figure}

In Fig.~\ref{fig:mass_reconstruction}, we show the ability of a sinjarite paleo-detector to constrain the DM mass from a future signal. The color scale shows contours in $\Delta m_\chi/m_\chi$; in the following we refer to the colored regions as those where the {\it mass can be reconstructed}. The gray regions indicate points in the parameter space where the $2\,\sigma$ confidence contours are not closed, i.e.~the reconstructed mass would be unbounded from above. Thus, we refer to the gray regions as portions of parameter space where {\it the mass cannot be reconstructed}. In Fig.~\ref{fig:mass_reconstruction}, we quantify how well the mass could be reconstructed for DM--nucleon cross sections between the projected 90\,\% confidence exclusion limits, cf. Sec.~\ref{sec:Optimistic}, and cross sections a factor 100 larger than this (the region bounded by the two black curves). Note that some of this region is already ruled out by XENON1T, cf. Fig.~\ref{fig:limits}.

Direct detection experiments suffer from an almost exact degeneracy between the mass and cross section at large DM masses \cite{Green:2008rd}. The degeneracy occurs when $m_{\chi}\gtrsim m_N$. This is because, for a given nuclear recoil energy, $v_{\mathrm{min}}$ depends only on the reduced mass of the DM--nucleus system. For $m_\chi \gg m_N$, the reduced mass becomes independent of the DM mass.

For traditional direct detection experiments, the ability to reconstruct the mass of a hypothetical DM particle has been studied extensively, see e.g.~\cite{Green:2007rb,Green:2008rd,Green:2008pf,Edwards:2018lsl,Bozorgnia:2018jep}. These studies found that a Xenon experiment would only be able to constrain the DM mass up to $\sim 200\,$GeV. For paleo-detectors, we instead find that mass reconstruction is possible for DM masses as large as $\sim 1\mathrm{\,TeV}$.

We show results for the high-resolution configuration in the upper panel of Fig.~\ref{fig:mass_reconstruction}. In the case of DM--nucleon cross sections just below the current limits, we find that the largest DM mass for which the mass could be reconstructed is $\sim 250\,$GeV. In the high exposure configuration (Fig.~\ref{fig:mass_reconstruction}, lower panel) the DM mass could be reconstructed for masses as high as $\sim 1\,$TeV if the DM--nucleon cross section is just below current limits. For cross sections further below current limits, we see that in the high resolution case the colored region extends to slightly larger masses than in the high exposure case.

 For a DM candidate with mass below $\sim 10\,$GeV, the mass could be reconstructed for all cross sections in reach of a sinjarite paleo-detector in either the high resolution or high exposure configuration. Although the precision of the mass reconstruction depreciates with decreasing DM--nucleon cross section, there are large regions of parameter space which are currently unconstrained and where the DM mass could be reconstructed reasonably well by paleo-detectors in case of a discovery.

The potential for paleo-detectors to tightly constraint the mass of a DM candidate stems in part from their large exposure. For example, in the high resolution case, a 6\,GeV DM candidate with cross section at the $90\,\%$ confidence exclusion limit would give rise to $\sim 10^5$ signal events. Such large numbers of events would allow us to accurately map out the track length spectrum. For DM masses of $\sim 1\,$TeV, paleo-detectors would only measure $\mathcal{O}(10^3)$ events at the exclusion limit, making the reconstruction of the associated track length spectrum more challenging. This should be contrasted with exposures possible in conventional direct detection experiments, where at their exclusion limits only $\mathcal{O}(1\text{-}10)$ events would be detected. Such a low number of signal events would not provide enough information to resolve minute differences in the recoil spectra required for mass reconstruction. 

Further, paleo-detectors could probe track lengths over three orders of magnitude, which corresponds to sensitivity spanning a large range of recoil energies. In particular, a $1000 \mathrm{\,nm}$ track  corresponds roughly to a $\sim1 \mathrm{\,MeV}$ nuclear recoil whereas a $1\mathrm{\,nm}$ track equates to a $\sim100\mathrm{\,eV}$ recoil\footnote{This can obviously depend significantly on the recoiling nucleus and target material.}. The high energy part of the recoil spectra has a significant dependence on the DM mass~\cite{Peter:2013aha,Bozorgnia:2018jep}. Unlike the traditional energy window of direct detection experiments, we exploit this information by observing a wide variety of track lengths. 

Finally, the target materials we consider here contain a variety of constituent nuclei with different masses. Sinjarite contains nuclei with masses from $1\,\mathrm{GeV}$ (H) to $37\,\mathrm{GeV}$ (Ca). Since the observed signal is a weighted sum over the contributions from the respective target nuclei, the resulting track length spectrum is richer in features than the simple slope dependence one finds for a single target nucleus. As these features can be exploited efficiently with a spectral analysis, paleo-detectors are particularly well suited for DM mass reconstruction.

We note that we have not considered any uncertainties in the DM velocity distribution in this analysis. For example, the longest tracks (which we leverage to constrain very heavy DM) are produced by the fastest moving DM particles, close to the Galactic escape velocity. The length of these longest tracks is therefore dependent on uncertainties in the escape velocity. More generally, allowing for variations in the DM distribution will widen the constraints on the DM mass. A number of techniques have been developed to incorporate velocity distribution uncertainties in direct detection (see for example Refs.~\cite{Strigari:2009zb,Fox:2010bz,Peter:2011eu,Kavanagh:2013wba,Feldstein:2014gza,Gelmini:2017aqe,Ibarra:2018yxq}) but we leave this more detailed analysis to future work. We note however that previous studies have shown that using multiple experiments with different target nuclei greatly reduces the impact of these uncertainties~\cite{Peter:2013aha,Edwards:2018lsl}. We expect the same to be true for paleo-detectors since the minerals investigated in this work contain a variety of target nuclei.

\section{Challenges}
\label{sec:Challenges}

Throughout this work we have shown that the sensitivity of paleo-detectors may well exceed that of current direct detection experiments. In Sec.~\ref{sec:Optimistic}, we projected the sensitivity assuming systematic errors on the overall normalization of the different background components only. For the neutrino-induced backgrounds, we assumed $100\,\%$ systematics, while we assumed $1\,\%$ systematics for backgrounds induced by radioactivity. In order to check the robustness of our results, we increased the normalization systematics on the neutron-induced backgrounds to 5\,\% and found that the sensitivities are unaffected. In the following, we discuss some of the other potential issues moving forward.

In our background normalization systematics case we assumed no covariance between the normalization of the background components. This assumption should hold for many background components, for example we expect no covariance between the spectra induced by solar neutrinos and diffuse supernova neutrinos. For the radiogenic backgrounds there may exist some covariance since they have a common origin. 

For the background shape systematics case, we assume no bin-to-bin covariances. This may be an optimistic or pessimistic assumption depending on the covariances one might expect. The most troublesome scenario would be a bin-to-bin covariance that makes signal-like variations in the background more likely. Due to the lack of theoretical guidance, we have chosen not to explore bin-to-bin covariances. Instead we attempt to maximize the error introduced by bin-to-bin systematics by using the minimum number of bins required to resolve all features in the spectra. We leave careful study of covariances to future analyses.

One of the primary assumptions throughout our analysis is that we can reject damage features in the minerals that are not tracks arising from nuclear recoils. Further, we do not consider the background produced by a series of linked $\alpha$-recoil tracks in the uranium-238 decay chain. The assumption is that the characteristic track pattern is easily recognizable and therefore rejected with 100\% efficiency. In reality there will be $>\mathcal{O}(10^7)$ tracks within a sample, many of which will exhibit these characteristic track patterns. We therefore require an automated tagging and rejection system. Since the characteristic track pattern is quite distinct from a normal track, it is possible that this system can be very efficient. However, this is yet to be validated for a large data set, a task we leave to future work.

The data produced by scanning significant amounts of material at high precision could present an issue by itself. Naively, scanning $1\mathrm{mm}^3$ of material at $1\mathrm{nm}$ precision will produce $10^6$ terabytes of data. It is a monumental task to analyze such a large data set. Luckily, much of the mineral will contain no track information at all, therefore a suitable compression format can be adopted to make the analysis more \href{https://www.merriam-webster.com/dictionary/tractable}{\color{black} tracktable}. Analysis of the data will require automated track-recognition, an ideal application of machine learning algorithms. We will also address this in future publications. 

Naively, one would expect paleo-detectors to be able to exploit directional information from the orientation of the recoil tracks. However, the target minerals we consider here are $\mathcal{O}(1\,\mathrm{Gyr})$-old, which is comparable to the period of the Sun's revolution around the Galactic center. Also, geological processes occur on timescales shorter than $\mathcal{O}(1\,\mathrm{Gyr})$, further complicating the expected directionality of the DM-induced signal. Reference~\cite{SnowdenIfft:1997hd} attempted to quantify the directional dependence of the DM-induced tracks within ancient minerals, showing that there is a preferred direction. Unfortunately for an $\mathcal{O}(1\,\mathrm{Gyr})$ mineral the effect in ancient mica was calculated to be only $\mathcal{O}(1\%)$. Because it is unlikely that we will be able to resolve the head/tail orientation of tracks at the nm scale, the induced anisotropy would need to be much larger than $\mathcal{O}(1\%)$ in order to be statistically observable~\cite{OHare:2017rag}. 

Finally, the translation of the range of the nucleus $x_T$ to the reconstructed track length after read-out is a source of uncertainty. Quantifying such an uncertainty requires detailed studies for different combinations of minerals and read-out methods~\cite{Baum:2018zzz}. However, in the case of a claimed detection, we would be able to confirm a signal using minerals with different constituent nuclei and ages, allowing one to mitigate some of these systematic issues.

\section{Conclusions}
\label{sec:Conclusion}

In this work, we have explored the prospects for probing Weakly Interacting Massive Particle (WIMP) Dark Matter (DM) with \textit{paleo-detectors}. In particular, we have extended previous studies by performing a full spectral analysis, including information about the expected distributions of track lengths left in the minerals by a DM signal as well as by neutrino-induced and radiogenic backgrounds. We further explored how systematic uncertainties on the normalization and shape of backgrounds impact projected limits. Finally, we have studied how well the DM mass could be measured in case of a future discovery.

We considered 4 minerals in this work: halite [\ch{NaCl}], olivine [\ch{Mg_{1.6}Fe^{2+}_{0.4}(SiO4)}], sinjarite [\ch{CaCl2$\cdot$ 2 (H2O)}] and nchwaningite [\ch{Mn^{2+}2SiO3(OH)2$\cdot$(H2O)}]. Sinjarite is the most sensitive out of the minerals examined here due to the assumed low levels of radioactive contamination and efficient neutron moderation by hydrogen (Ref.~\cite{Baum:2018zzz} came to a similar conclusion for the mineral epsomite [\ch{Mg(SO4)$\cdot$ 7 (H2O)}]). 

For moderate track length resolutions, $\sigma_{x_T} = 15\,\mathrm{nm}$, we find that the full spectral analysis extends the projected paleo-detector sensitivity to DM--nucleon cross sections roughly an order of magnitude smaller than the sliding window analysis of Refs.~\cite{Baum:2018tfw,Baum:2018zzz}. This improvement is driven by the fact that a full spectral analysis automatically entails the use of optimal `control regions', where the signal is sub-dominant, helping to pin down the normalization and shape of the backgrounds. 

We find that by analyzing an $\mathcal{O}(10 \,\mathrm{cm}^3)$ sample of sinjarite using small angle X-ray scattering, it could be possible to probe DM--nucleon cross sections roughly a factor of 100 smaller than current direct detection experiments for DM heavier than $\sim 100\,$GeV. Including systematic uncertainties in background shapes at the 10\,\%-level, projected limits remain a factor of $7-8$  more stringent than current XENON1T bounds~\cite{Aprile:2018dbl}. The sensitivity depreciates further if systematics larger than 10\,\% are assumed for the shapes of backgrounds.

Analyzing smaller samples of $\mathcal{O}(1\,\mathrm{mm}^3)$ at nm-resolution (e.g.~using helium ion beam spectroscopy), we find that paleo-detectors may be able to probe DM--nucleon scattering cross sections many orders of magnitude below current limits, for $500\,\mathrm{MeV} \lesssim m_\mathrm{\chi} \lesssim 10\,\mathrm{GeV}$. Probing $\mathcal{O}(\mathrm{nm})$ track lengths corresponds to an $\mathcal{O}(100\,\mathrm{eV})$ energy threshold, exploring significant regions of the recoil spectra from low-mass WIMPs. With high-resolution read-out methods, the limits would be robust to systematic uncertainties in the background shapes as large as $\sim 50\,\%$.
 
In addition,  we have investigated the prospects for paleo-detectors to constrain the DM parameters in the case of a future signal. As an example, we calculate the regions in which the mass and cross section become degenerate for a sinjarite paleo-detector. We find that below $m_\chi \lesssim 15\mathrm{\,GeV}$, it would be possible to reconstruct the mass of the DM particle with a relative error of less than 10\,\% if the cross section is large enough for a $5\,\sigma$ discovery. For $m_\chi \gtrsim 15\mathrm{\,GeV}$ the signal becomes increasingly insensitive to changes in $m_{\chi}$, making the mass harder to constrain. In spite of this, paleo-detectors should be able to obtain both a lower and an upper limit on the DM mass for $m_\chi \lesssim 1\,\mathrm{TeV}$ if the cross section is just below current limits. In contrast, conventional direct detection experiments could provide only a lower bound on the DM mass if the true mass is larger than $\sim 200$\,GeV~\cite{Edwards:2018lsl}.

Paleo-detectors could also be used to investigate a number of interesting questions beyond searches for DM. For example, in the absence of a DM signal, the analysis outlined here would straightforwardly allow us to measure neutrino-induced events. It could therefore be possible to use mineral samples of different ages as a unique probe of the neutrino history of our galaxy. This will potentially allow us to study both historical neutrino processes in the Sun and the signal from supernovae. 

Paleo-detectors represent an excellent opportunity to probe large areas of the WIMP DM parameter space in the near future. The next steps involve assessing the practical challenges of reaching the required exposures to achieve these sensitivities, as well as more detailed modeling of backgrounds. We leave both of these tasks to future work. However, we note that WIMP-nucleon cross sections much smaller than projected in this work may be probed by paleo-detectors if either novel ideas to control the backgrounds emerge (akin to progress made in conventional direct detection experiments in recent decades) or if target materials with significantly lower levels of radioactivity are available. With uranium concentrations of $\lesssim 10^{-15}$, radioactive backgrounds would no longer dominate at high DM masses. In such a case, paleo detectors could perhaps probe WIMP-nucleon cross sections all the way down to the diffuse supernova and atmospheric neutrino floor.

\acknowledgements
We thank Niki Klop and Adri Duivenvoorden for bringing this wonderful collaboration together. 
SB would like to thank GRAPPA and the University of Amsterdam, where part of this work was completed, for hospitality. 
This research is partially funded by NWO through the VIDI research program ``Probing the Genesis of Dark Matter" (680-47-532; TE, BK, CW). 
SB, AKD, KF, and PS acknowledge support by the Vetenskapsr\r{a}det (Swedish Research Council) through contract No. 638-2013-8993 and the Oskar Klein Centre for Cosmoparticle Physics. 
SB, KF, and PS also acknowledge support from DoE grant DE-SC007859 and the LCTP at the University of Michigan. 
We also greatly thank the makers of WINE, for making it possible for us to run SRIM. 
Finally, we acknowledge the use of a number of packages for scientific computing in Python~\cite{Oliphant2007,Scipy1,Hunter2007,Perez2007,Millman2011,Oliphant:2015:GN:2886196,scikit-learn}.

\appendix

\section{Euclideanized Signals}\label{sec:EuclMeth}

Whether an experiment is \textit{a priori} able to constrain a parameter of interest involves calculating the \textit{expected statistical distinctness} between two signals, given a set of backgrounds and their associated uncertainties \cite{Algeri:2018zph}. Points in the model parameter space are described by a $d$-dimensional vector $\boldsymbol\theta = (\theta_1,\theta_2,\dots,\theta_d)$. Two model parameter points $\boldsymbol\theta^{(1)}, \,\boldsymbol\theta^{(2)}$ can be considered as experimentally distinguishable if the parameter point $\boldsymbol\theta^{(2)}$ is inconsistent (at a given significance level) with the Asimov data $\mathcal{D} = \DA(\boldsymbol\theta^{(1)})$. For our application we have a two dimensional model where $\boldsymbol\theta^{(i)} = \{ m_\chi^{(i)}, (\sigma_n^{\rm SI})^{(i)}\}$.
In order to establish experimental distinguishability, we use the maximum-likelihood ratio as a test statistic (TS)~\cite{Wilks:1938a, Cowan:2010js},
\begin{equation}
  \text{TS}(\boldsymbol\theta^{(2)},\boldsymbol\theta^{(1)}) \equiv- 2 \ln
  \frac
  {{\max_{\boldsymbol \eta}}\mathcal{L}
  (\DA(\boldsymbol\theta^{(1)})|\boldsymbol\theta^{(2)}, \boldsymbol\eta)}
  {{\max_{\boldsymbol\eta}}\mathcal{L}
  (\DA(\boldsymbol\theta^{(1)})|\boldsymbol\theta^{(1)}, \boldsymbol\eta)}\;,
  \label{eqn:TS}
\end{equation}
where $\mathcal{L}(\mathcal{D}|\boldsymbol\theta, \boldsymbol\eta)$ is the likelihood function for data $\mathcal{D}$.  It can depend on some nuisance parameters $\boldsymbol\eta$ that are profiled out when calculating TS.
For model parameter points with sufficiently similar signals, the value of TS is approximately symmetric under $\boldsymbol\theta^{(1)} \leftrightarrow \boldsymbol\theta^{(2)}$.
Hence, we can write
\begin{equation}
    \label{eqn:fisher}
    \text{TS}(\boldsymbol\theta^{(2)}, \boldsymbol\theta^{(1)}) \approx (\boldsymbol\theta^{(1)}-\boldsymbol\theta^{(2)})^{T}\mathcal{I}(\boldsymbol\theta^{(1)}-\boldsymbol\theta^{(2)})\;,
\end{equation}
where
\begin{equation}
\mathcal{I}_{ij} = -\left\langle \frac{\partial^2\log\mathcal{L}(\mathcal{D}|\boldsymbol\theta^{(1)})}
{\partial\theta_i\partial\theta_j}\right\rangle_{\mathcal{D}(\boldsymbol\theta^{(1)})}
\;,
\end{equation}
is the Fisher information matrix at $\boldsymbol\theta^{(1)}$. The derivatives here describe the curvature in the direction of a particular parameter. The Fisher information matrix defines a metric on the space of model parameters, making it accessible to the tools of differential geometry.

\medskip

The \textit{Euclideanized signal method} is an approximate isometric embedding of a $d$-dimensional model parameter space (with geometry from the Fisher information metric) into $n$-dimensional Euclidean space:
$\vect \theta \mapsto \vect x(\vect \theta)$ with $\vect x\in \mathcal{M} \subset\mathbb{R}^n$ and $\vect \theta \in \mathbb{R}^d$.
This embedding allows one to estimate differences in the log-likelihood ratio by the Euclidean distance,
\begin{equation}
\mathrm{TS}(\boldsymbol\theta^{(2)}, \boldsymbol\theta^{(1)}) 
\simeq \lVert (\vect x(\boldsymbol\theta^{(1)}) - \vect x(\boldsymbol\theta^{(2)})\rVert^2\;.
\end{equation}
Machine learning tools (in particular clustering algorithms, which usually assume Euclidean space) can then be used to efficiently explore the signal phenomenology of different models, and to systematically compare entire model classes, see Ref.~\cite{scikit-learn}. For details on the Euclideanized signal transformation and its accuracy see Ref.~\cite{Edwards:2017kqw}. The accuracy of the method (relative to the TS value) is at the $< 20\%$ level, and details can be found in Ref.~\cite{Edwards:2018lsl}.

We can now estimate the ability of an experiment to constrain the mass in the following way:
\begin{itemize}
    \item Grid scan the parameter space, calculating signals for each point $\boldsymbol\theta^{(i)}$.  Here, it is essential that all distinguishable model parameter points are covered down to a specific significance level (this should correspond to approximately 10 points per $1\,\sigma$ region).
    \item Euclideanize the signals using \texttt{swordfish} to produce associated vectors $\mathbf{x}_i$. Note that the transformation is able to account for arbitrary Gaussian background uncertainties and correlations, see Refs.~\cite{Edwards:2018lsl,Edwards:2017kqw} for more details.
    \item For each parameter point $\boldsymbol\theta^{(i)}$ we calculate its associated nearest neighbours within a predefined significance. Here we use $2\,\sigma$ which corresponds to a radius of $\sqrt{2}$ in one dimension. The number of dimensions reflects the difference in dimensionality between the two-parameter model with \{$m_{\chi},\sigma^{\mathrm{SI}}_n$\} and the model living on the high mass boundary where $m_{\chi}$ is fixed. If this set of nearest neighbours contains a parameter point on the high mass boundary\footnote{Here we define the high mass boundary as $10\mathrm{\,TeV}$.} the constraint on the mass around $\boldsymbol\theta^{(i)}$ is unbounded from above.
\end{itemize}

\bibliographystyle{apsrev4-1}
\bibliography{refs}

\end{document}